\documentclass[manuscript,nonacm]{acmart}
\newcommand{\proj}[0]{Conflect}
\usepackage{booktabs}
\usepackage{graphicx}
\usepackage{tikz}
\usepackage{tabularx}
\usepackage{multirow}
\usepackage{multicol}
\usepackage[xcdraw,table]{xcolor}
\usetikzlibrary{trees}
\usepackage{subfig}
\AtBeginDocument{%
  \providecommand\BibTeX{{%
    \normalfont B\kern-0.5em{\scshape i\kern-0.25em b}\kern-0.8em\TeX}}}

\setcopyright{acmcopyright}
\copyrightyear{2018}
\acmYear{2018}
\acmDOI{XXXXXXX.XXXXXXX}

\acmConference[Conference acronym 'XX]{Make sure to enter the correct
  conference title from your rights confirmation emai}{June 03--05,
  2018}{Woodstock, NY}
\acmPrice{15.00}
\acmISBN{978-1-4503-XXXX-X/18/06}

\begin{document}

\title[\proj{}: Designing Reflective Thinking-Based\\ Contextual Privacy Policy for Mobile Applications]{\proj{}: Designing Reflective Thinking-Based Contextual Privacy Policy for Mobile Applications}

\author{Shuning Zhang}
\authornotemark[1]
\orcid{0000-0002-4145-117X}
\affiliation{%
  \institution{Tsinghua University}
  \city{Beijing}
  \country{China}
}
\email{zsn23@mails.tsinghua.edu.cn}

\author{Sixing Tao}
\authornote{Equal contribution.}
\affiliation{
    \institution{University of Washington}
    \city{Seattle}
    \country{United States}
}
\email{sixint@uw.edu}

\author{Eve He}
\affiliation{
    \institution{Independent Researcher}
    \city{Madison}
    \country{United States}
}
\email{eveanny.hx@gmail.com}

\author{Yuting Yang}
\affiliation{
    \institution{Independent Researcher}
    \city{Ann Arbor}
    \country{United States}
}
\email{yutingyang986@gmail.com}

\author{Ying Ma}
\affiliation{
    \institution{School of Computing and Information Systems, The University of Melbourne}
    \city{Melbourne}
    \country{Australia}
}

\author{Ailei Wang}
\affiliation{
    \institution{Academy of Arts \& Design, Tsinghua University}
    \city{Beijing}
    \country{China}
}

\author{Xin Yi}
\orcid{0000-0001-8041-7962}
\authornote{Corresponding author.}
\affiliation{
    \institution{Tsinghua University}
    \city{Beijing}
    \country{China}
}
\email{yixin@tsinghua.edu.cn}

\author{Hewu Li}
\orcid{0000-0002-6331-6542}
\affiliation{
    \institution{Tsinghua University}
    \city{Beijing}
    \country{China}
}
\email{lihewu@cernet.edu.cn}

\renewcommand{\shortauthors}{Trovato and Tobin, et al.}


\begin{abstract}
    % Privacy policies are often lengthy and complex, leading to user neglect. While contextual privacy policies (CPPs) aim to present information at the point of risk, they often lack engagement and can disrupt user tasks. We introduce Conflect, a interactive CPP prototype for mobile applications guided by a reflective thinking framework. We first conducted three workshops with experienced designers and researchers to identify key problems and designs. Participants ranked the disconnect between context and action as the most critical issue. Guided by participants' highest-rated designs, we designed and implemented Conflect, which hints users with sidebar alerts, enabling them to reflect on potential risks and take just-in-time control. Conflect detected privacy risks contextually, extracted and matched correlated policy segments and generated risk descriptions automatically. Technical evaluations demonstrated Conflect's feasibility in extracting privacy policies with 94.0\% accuracy on the CPP4APP dataset and an end-to-end latency of 4.35\,s. The subsequent user study (N=28) comparing Conflect with traditional CPP, privacy policy, and privacy labels confirms Conflect's effectiveness in improving user understanding, trust, and satisfaction while lowering cognitive load.
    Privacy policies are lengthy and complex, leading to user neglect. While contextual privacy policies (CPPs) present information at the point of risk, they may lack engagement and disrupt tasks. We propose Conflect, an interactive CPP for mobile apps, guided by a reflective thinking framework. Through three workshops with experienced designers and researchers, we constructed the design space of reflective thinking-based CPP design, and identified the disconnect between context and action as the most critical problem. Based on participants' feedback, we designed Conflect to use sidebar alerts, allowing users to reflect on contextualized risks and fostering their control. Our system contextually detects privacy risks, extracts policy segments, and automatically generates risk descriptions with 94.0\% policy extraction accuracy on CPP4APP dataset and a 4.35s latency. A user study (N=28) demonstrated that Conflect improves user understanding, trust, and satisfaction while lowering cognitive load compared to CPPs, privacy policies and privacy labels.
\end{abstract}




\begin{teaserfigure}
    \centering
    \subfloat[]{
        \includegraphics[width=0.19\textwidth]{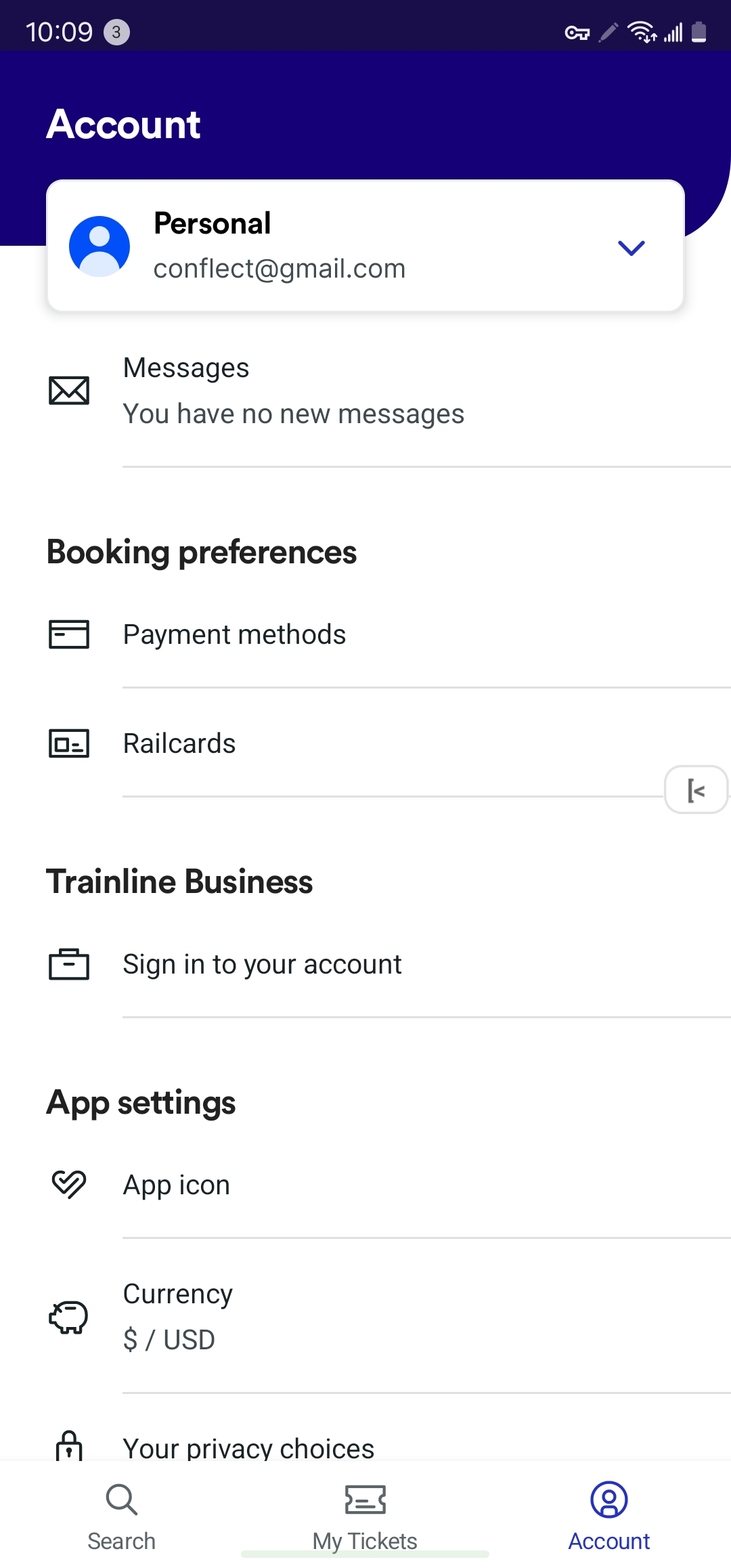}
        \label{fig:teaser_1}
    }
    \subfloat[]{
        \includegraphics[width=0.19\textwidth]{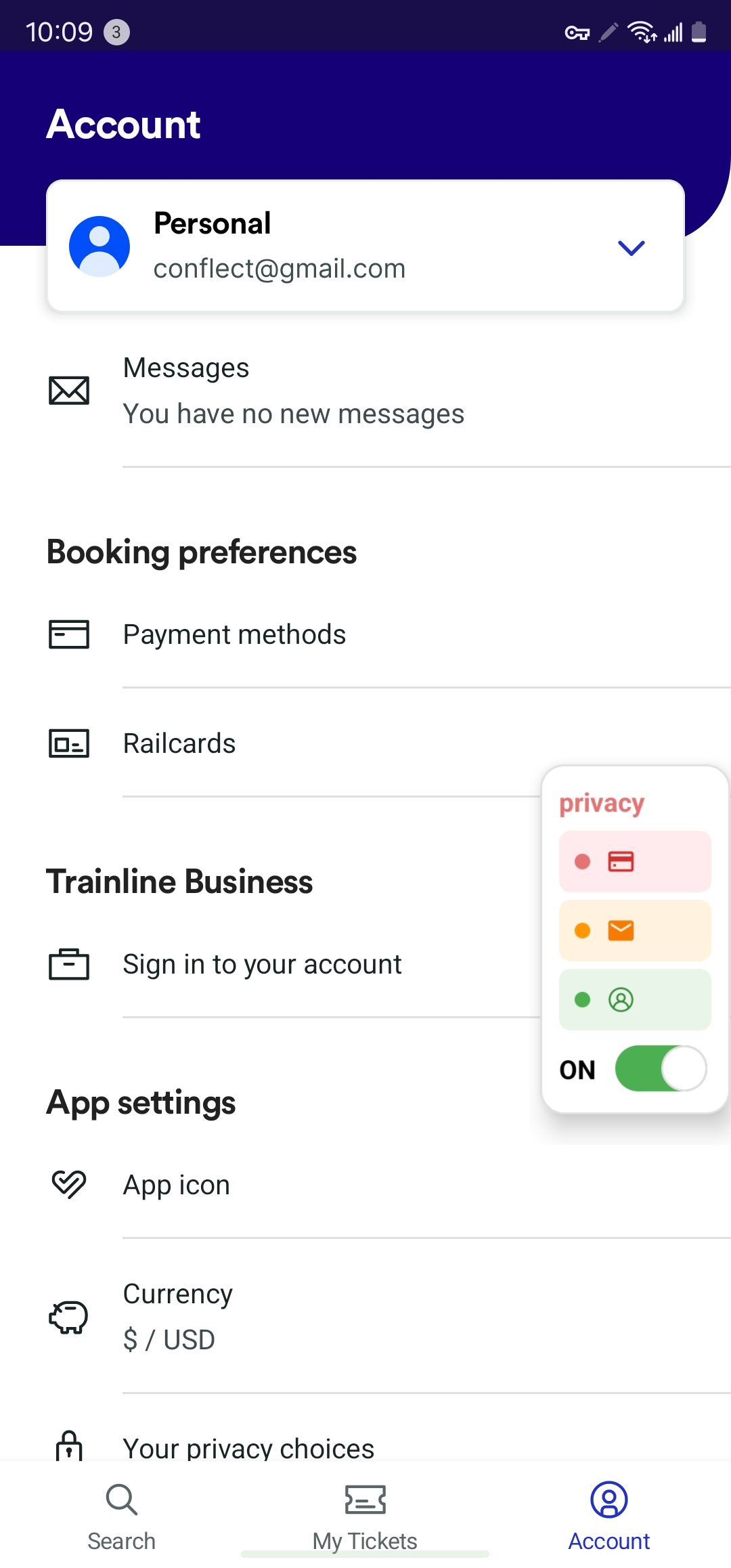}
        \label{fig:teaser_2}
    }
    \subfloat[]{
        \includegraphics[width=0.19\textwidth]{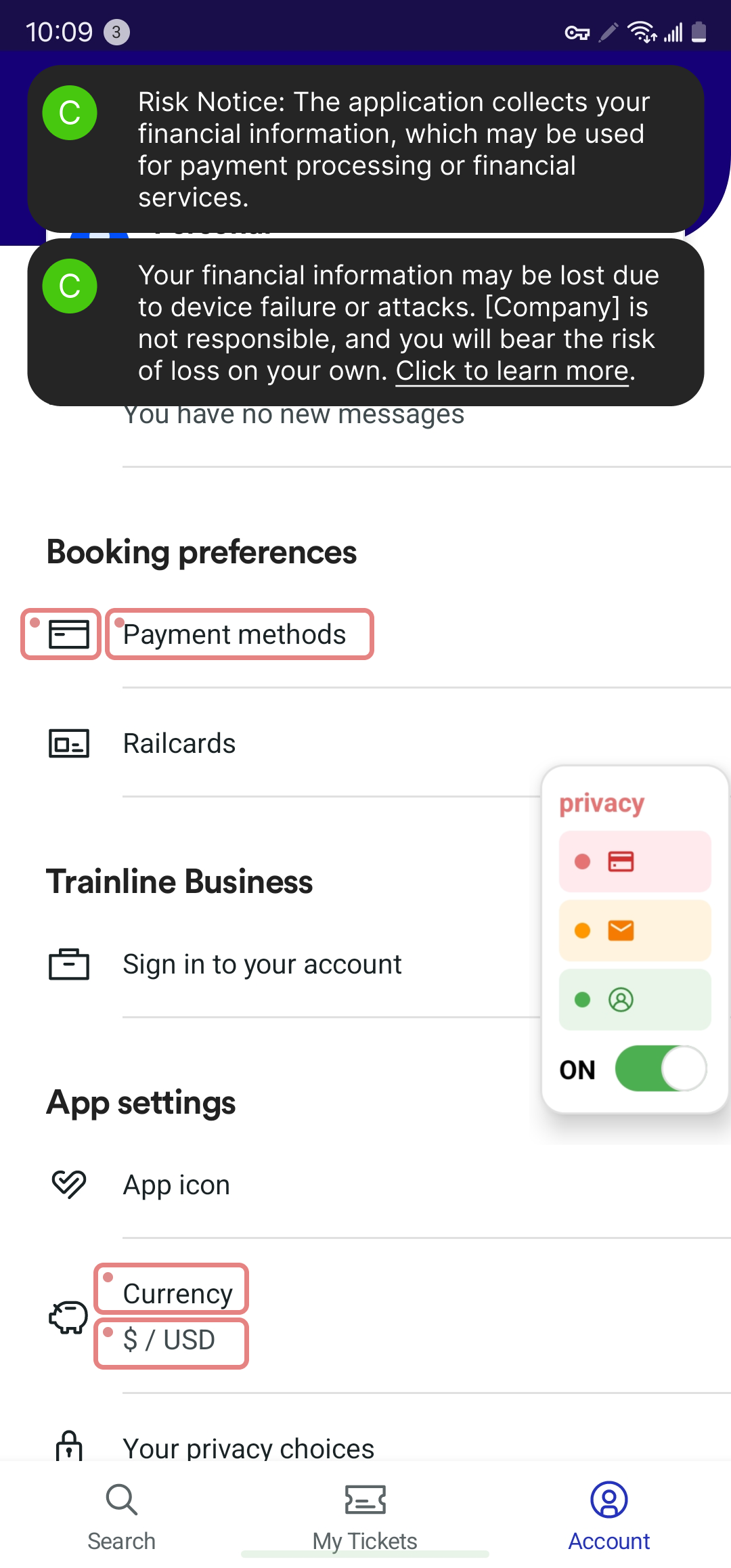}
        \label{fig:teaser_3}
    }
    \subfloat[]{
        \includegraphics[width=0.19\textwidth]{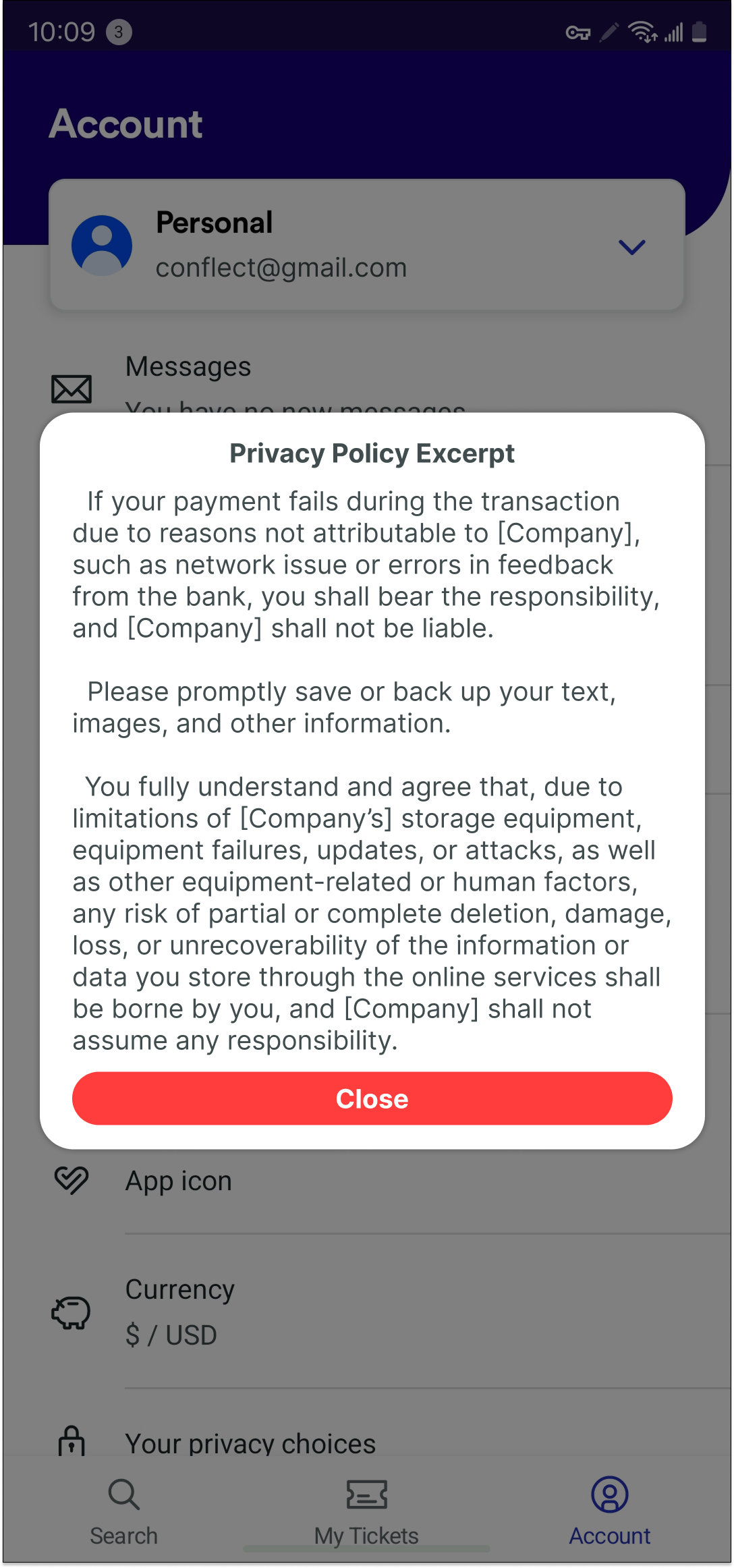}
        \label{fig:teaser_4}
    }
    \subfloat[]{
        \includegraphics[width=0.19\textwidth]{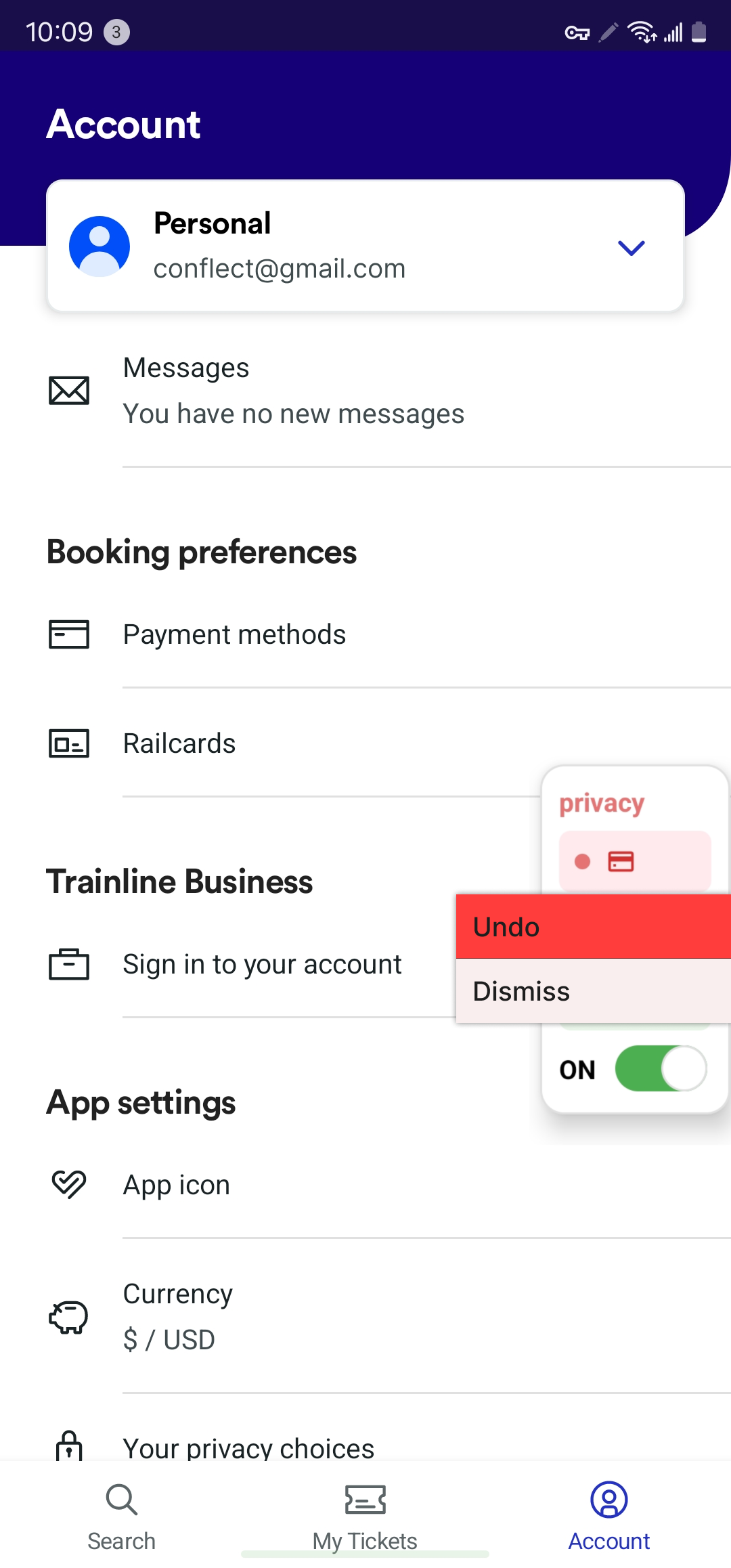}
        \label{fig:teaser_5}
    }
    \caption{Illustration of \proj{}, a reflective-thinking-based contextual privacy policy prototype. Unlike traditional privacy policies that appear once before app usage, \proj{} dynamically surfaces context-specific risks during interaction. (a) When collapsed, \proj{} minimizes into a small arrow at the right screen edge (as illustrated here in the middle), which jolts leftward with a red border to notify users when new privacy risks emerge. (b) Tapping the collapsed arrow expands it into a floating window that displays detected privacy risks with corresponding severity indicators (color-coded). These \textit{served as the experience stage}. (c) Tapping a specific risk icon simultaneously triggers bounding boxes and the first pop-up, which highlights privacy data collection; the bounding boxes remain visible as the second pop-up appears, presenting scenario-based potential risks. (d) Tapping the second pop-up further reveals the corresponding privacy policy excerpt. These \textit{served as the reflection stage}. (e) By default, when users encounter the same risk trigger again, the alert will reappear to remind them. A long-press on the alert enables users to permanently dismiss the highlighted risk triggers. Users can also directly take action within the host application to adjust privacy permissions, These constituted just-in-time control and \textit{served as the action stage}.}
    % The interaction of \proj{} adopting a reflective-thinking based framework, (a) where it uses alerts that are color-coded according to their sensitivity, on the sidebar to hint users of potential privacy collection and risks, as well as highlighting privacy content on the interface using bounding boxes, acting as the \textit{experience} stage. When users click on the alerts, the notifications pops into the interface, hinting users about potential risks, acting as the \textit{reflection} stage. These notifications hint users to control the privacy, where users could also clicked on the notification to (b) see the original privacy policy segment. A long-click on the alert (c) prompted users for just-in-time binary control, where they could just ignore the risk or control the risk.
    \label{fig:teaser}
\end{teaserfigure}


\maketitle

\section{Introduction}

Privacy policies serve as the primary mechanism for informing users about data practices in mobile applications~\cite{adams2020agreeing,flavian2006consumer}. However, their effectiveness is significantly undermined by their excessive length and complexity. Research indicates that the average privacy policy contains approximately 4,000 words, requiring an estimated 16 minutes for an average adult to read~\cite{blakkarly2022privacy}. Hindered by these aspects, 74\% of users were reported to bypass these policies entirely, opting instead for the ``quick join'' option without engaging with the content~\cite{obar2020biggest}. This behavior is largely attributed to the cognitive burden imposed by the verbose and legalistic nature of traditional privacy notices. Meanwhile, traditional privacy policies are typically presented only once at the beginning of app use, meaning that users may not be aware of relevant privacy risks later in the interaction~\cite{pan2024new}.

In response to these limitations, the concept of ``just-in-time'' privacy notices has emerged as a more user-centric alternative~\cite{schaub2015design}. These notices aim to present privacy information at the moment of relevance. However, existing implementations suffer from several shortcomings: (1) it is primarily limited to the installation phase of applications~\cite{scoccia2018investigation}, (2) they offer only a limited snapshot of data access, without conveying the broader context of how data maybe transmitted, shared, stored or repurposed, and (3) they often lack contextual integration, resulting in low user engagement and limited effectiveness~\cite{felt2010effectiveness, ma2014android}.

To address these issues, researchers have proposed Contextual Privacy Policies (CPPs)~\cite{windl2022automating,ortloff2020implementation}, which decompose traditional privacy policies into concise, contextually relevant segments. CPPs are designed to deliver information at the right time and at a manageable rate, thereby reducing cognitive overload and improving comprehension~\cite{patil2015interrupt,windl2022automating}. Recent work has explored automated methods for generating such policies~\cite{windl2022automating,pan2024new}, offering promising directions for scalable implementation.

Despite these advances, current CPPs remain limited in their interactive capabilities, and can be disruptive~\cite{pan2024new}. While more engaging modalities like serious games can improve comprehension~\cite{stellmacher2022escaping}, they are often too intrusive for everyday mobile contexts. This presents a central design challenge: \textbf{how to create a CPP that is both engaging and minimally intrusive.} To address this, we propose grounding the design of CPPs in reflective thinking framework~\cite{jasper2003beginning,driscoll2006practising}. This approach fosters engagement not through overt interruption, but by encouraging users to connect abstract policy information to their concrete actions and potential consequences. By transforming passive interruption into active deliberation, this framework offers a promising path for enhancing the effectiveness of CPPs. To explore this potential, we seek to answer the following research questions (RQs):
% An alternative framework is reflective thinking~\cite{jasper2003beginning,driscoll2006practising}. This approach, draw from educational history, transforms passive notification into active deliberation by encouraging users to pause and connect abstract policy text to their concrete actions and potential consequences. To address this gap, the present study seeks to enhance the effectiveness and user engagement of CPPs by answering the following research questions (RQs): 

% formative study
\textbf{RQ1:} What are the design opportunities of CPPs on mobile devices guided by reflective thinking framework?

\textbf{RQ2:} How can we design a non-intrusive CPP that delivers timely privacy information without disrupting the user's primary task?

\textbf{RQ3:} How does a CPP designed with a reflective thinking framework impact users' understanding, cognitive load, and behavior compared to traditional and other CPP formats?

To address RQ1 and inspire the design of \proj{}, we first conducted three workshops with experienced designers and researchers. Through a multi-stage design process, participants identified the \textit{disconnect between abstract policy text and concrete user actions} as the most critical problem for CPPs to solve. Consequently, the design patterns that participants prioritized were those that directly bridged this gap across the stages of reflective thinking. For the initial experience stage, they favored non-intrusive methods like \textit{ambient and minimalist cues}. For the reflection stage, there was a strong consensus on translating policy into \textit{contextualized, risk-based scenarios}, supplemented by \textit{summarization and layered disclosure}, to make abstract risks tangible. Finally, for the action stage, \textit{just-in-time granular controls} were rated highest, underscoring the importance of empowering users with immediate action choices.

Informed by these findings, we designed and implemented \proj{}, a CPP that operationalizes this reflective thinking framework (Figure~\ref{fig:teaser}). By default, it manifests as a minimal sidebar that, when interacted with, highlights relevant UI elements to ground the user's experience in the immediate context. It then presents notifications that frame data practices as tangible, risk-based scenarios, and allows user to view the corresponding policy excerpts to facilitate reflection. Finally, it nudges users toward an actionable choice, such as adjusting privacy permissions within the host application or dismissing or ignoring recurring alerts, thereby completing the action stage. 

Technically, \proj{}'s algorithm comprises three modules, which adapts the initial stages of on-screen element detection and policy extraction from Pan et al.~\cite{pan2024new}. We additionally designed a reflective presentation module that leverages a Large Language Model (LLM) to synthesize the legal text into concise, context-aware descriptions of data practices and potential risks. A technical evaluation validated our implementation, which achieved 94.0\% accuracy for policy extraction, an average satisfaction score of 6.4 out of 7 for its generated risk descriptions on CPP4APP dataset~\cite{pan2024new}. Furthermore, \proj{} achieves an average latency of 4.35\,s (SD 0.93\,s) from screenshot arrival to scenario display on a server with 8~vCPUs and 32~GB RAM.
% , representing a 78\% reduction compared to the serial baseline.
% with the first two modules following Pan et al.'s practice~\cite{pan2024new} and the third module highlighting contextual risk generation. First, a contextual privacy detection module employs PaddleOCR and ResNet to identify privacy-salient text and icons on the user's screen. Subsequently, a privacy policy extraction module automatically retrieves and parses the application's privacy, utilizing a LLM to extract structured data practices. Finally, a novel reflective presentation module maps the detected on-screen elements to relevant policy segments, and leverages LLMs to synthesize the legal text into concise, context-aware data practice descriptions, and risk scenarios. A technical evaluation demonstrated \proj{}'s effectiveness: \proj{} achieved 94.0\% accuracy in privacy policy extraction, an average satisfaction score of 6.4 out of 7 for its generated risk descriptions, and an end-to-end latency of XX.X seconds on XX device.

To evaluate the impact of reflective thinking-based CPPs (RQ3), we conducted a user study (N=28) comparing \proj{} against traditional privacy policy, privacy label~\cite{kelley2009nutrition}, and a start-of-the-art CPP for mobile applications~\cite{pan2024new}. The results showed that \proj{} significantly lowered cognitive load while improving usability and user experience scores compared to baseline methods. While no significant effect on policy understanding was found, qualitative feedback indicated that \proj{} increased participants' awareness of otherwise unnoticed data collection practices. Participants also praised its non-intrusive nature and the enhanced sense of control it provided. In summary, this paper makes three contributions:

$\bullet$ We articulate a novel design space for CPPs based on reflective thinking framework to guide the development of CPPs.

$\bullet$ We design and implement \proj{}, a technique that engages users while reducing their cognitive load.

$\bullet$ We empirically demonstrate that our reflective approach enhances usability and engagement compared to traditional and existing contextual policy designs.

\section{Background and Related Work}

We first introduced the landscape of privacy policy visualization, then introduced the development of contextual privacy policy. We finally synthesized aspects around reflective thinking-based design.

\subsection{Privacy Policy Visualization}

Initial efforts in privacy policy visualization centerd on machine-readable formats like the Platform for Privacy Preferences (P3P) and tools such as PrivacyBird~\cite{cranor2002web}, which aimed to be both human-understandable and machine-readable. However, achieving widespread adoption proved challenging. Consequently, research shifted toward condensing policy information into digestible formats, such as privacy nutrition labels~\cite{kelley2009nutrition,kelley2010standardizing} and related tabular forms~\cite{reeder2008user,kelley2010standardizing}. While these formats were later enhanced with interactivity~\cite{reinhardt2021visual} and adapted for mobile applications~\cite{zhang2025privcaptcha}, a key limitation emained that the visualizations were not presented at the precise moment a privacy risk occurred. This temporal disconnect limited their effectiveness in informed consent.

A parallel stream of research explored expressive and engaging modalities, including comics~\cite{tabassum2018increasing}, and serious games~\cite{stellmacher2022escaping}. Although these approaches successfully increased user engagement, they introduced practical challenges, such as lengthy interaction times and difficulties in automatic generation.

Finally, embedding privacy notices in a contextual form has emerged as a promising approach. This method is effective because it directly correlates the visualization with the specific privacy risk at hand and condenses complex information into an easily understandable format~\cite{pan2024new}. This line of work has progressed from initial conceptualizations~\cite{ortloff2020implementation} to practical implementations on websites~\cite{windl2022automating} and mobile apps~\cite{pan2024new}. However, existing contextual notices still lack the interface design to promote user reflection on privacy risks, particularly on screen-constrained mobile devices. Addressing this gap, this paper adopts a reflective thinking perspective to design and visualize contextual privacy policies for mobile applications.

\subsection{Contextual Privacy Policy}

Contextual Privacy Policies (CPPs) present privacy information within relevant user interaction contexts. Early research established the conceptual foundations for CPPs by challenging the monolithic structure of traditional privacy policies. Bolchini et al.~\cite{bolchini2004need} first proposed reorganizing policies around specific user interaction contexts, while Feth et al.~\cite{feth2017transparency} refined this by advocating for contextual statements tailored to users' immediate actions. However, these foundational works focused on information architecture and did not explore interactive designs to promote user engagement with the contextualized context.

% CPPs presents privacy policy information within relevant user interaction contexts. Early research established the foundation for CPPs. Bolchini et al.~\cite{bolchini2004need} identified the limitations of traditional privacy policies as monolithic documents that impeded users from accessing essential privacy details. They introduced a method to reorganize privacy policies around specific user interaction contexts. Building on this, Feth et al.~\cite{feth2017transparency} further developed the concept by proposing contextual privacy statements that align privacy information more closely with user immediate contexts, moving away from generic one-size-fits-all policies.

Subsequent research validated the effectiveness of CPPs. Patil et al.~\cite{patil2015interrupt} highlighted that privacy feedback is most effective when delivered at the right time and at a manageable rate, both of which CPPs facilitate by reducing the amount of text displayed at once and presenting information contextually. Bergmann et al.~\cite{bergmann2008testing} showed that displaying privacy information in corresponding contexts significantly increases users' privacy awareness. Empirical studies showed that CPPs enhance privacy notice transparency, making a progressive step toward more user-centric privacy communication~\cite{masotina2022transparency,ortloff2020implementation}. Ortloff et al.~\cite{ortloff2020implementation} evaluated CPPs using a concept showcase for seven websites, manually identifying suitable placements and extracting privacy policy information. While these studies confirmed the benefits of contextual information delivery, they did not investigate how the \textit{design} of the CPP itself could encourage deep user reflection on potential privacy risks.

Recent advancements have focused on automating CPPs' generation. Ortloff's manual approach~\cite{ortloff2020implementation} lacks scalability given the millions of websites in existence. Previous research has also developed methods for automatic privacy policy processing. Zimmeck and Bellovin~\cite{zimmeck2014privee} introduced Privee, an architecture for automatic privacy policy analysis, while Mysore Sathyendra et al.~\cite{sathyendra2017identifying} developed classification models to identify opt-out choices in policies. Harkous et al.~\cite{harkous2018polisis} introduced Polisis, an automated framework using hierarchical neural network classifiers for high-level and fine-grained queries, offering a promising approach for annotating entire policies.

More recent advancements have produced end-to-end tools capable of automatically generating and injecting contextual notices. Windl et al.~\cite{windl2022automating} proposed PrivacyInjector, a production AI tool designed to automatically generate contextual privacy policies for websites. Nevertheless, its data practice identification is limited to six website-specific types. Pan et al.~\cite{pan2024new} extended CPPs to mobile applications, presenting a novel framework for generating CPPs in that context. Despite these technical achievements, existing implementations still largely position users as passive information recipients, and the design of CPPs to demonstrate privacy policy is important because it greatly affected users' information acquiring, but under-studied.

\subsection{Reflective Thinking-based Design}

Reflective practice, a concept with origins in educational disciplines, facilitates a deep analysis of situations by challenging assumptions and fostering new perspectives~\cite{jasper2003beginning,kolb1984experiential}. While existing research has advanced privacy policy visualization and CPPs~\cite{eikey2021beyond,marougkas2023virtual}, most approaches focus on information transfer and largely position users as passive recipients of information. They often lack mechanisms for deep engagement, thereby limiting users' capacity to critically assess personal privacy risks. Reflection, in contrast, empowers users to pause, evaluate assumptions and connect abstract risks to personal contexts. Adopting reflective practice from fields such as education and design~\cite{schon1983reflective,sengers2005reflective}, this work leverages it as a foundation to address the gap in CPP design. By fostering reflectivg engagement, CPPs can evolve from static notices into interactive tools that empower user agency. 

Several models have been developed to structure this reflective process~\cite{bentvelzen2022revisiting}. The Experience-Reflection-Action (ERA) cycle is a simple, cyclical framework that encourages users to recall an experience, reflect on its meaning, and act on new insights~\cite{jasper2003beginning}. Similarly, Driscoll's Framework, structured around the questions ``What?'', ``So what?'', and ``Now what?'', offers a clear, adaptable approach for contexts like CPPs where users must quickly interpret information and make decisions~\cite{driscoll1994reflective,driscoll2006practising}. In comparison, more elaborate frameworks such as Kolb's four-stage experiential learning cycle and Gibb's comprehensive six-stage reflective cycle offer depth but may be less practical for immediate privacy decisions, as they require extended engagement that can overwhelm users in fast-paced mobile environments \cite{kolb1984experiential,gibbs1988learning,padiyath2025critical,pan2024new,li2022cultural}. The contrast between these models highlights the importance of balancing accessibility with richness. This work adopts the lightweight structures of the ERA and Driscoll models due to their pragmatic suitability for designing CPPs that are encountered in time-constrained situations \cite{driscoll1994reflective,driscoll2006practising}.

\section{Formative Study: Reflective Thinking Framework-Based Workshops For CPP Design}

% Although there were many papers focusing on delineating the problems of privacy policies~\cite{zhang2025privcaptcha,reinhardt2021visual}, to ground the problems of privacy policies that could be solved by CPPs, and answer RQ1, we conducted workshops. These workshops involved design and privacy \& security experts to brainstorm potential interaction designs, and converge to a final satisfying candidate.
To address RQ1, we conducted formative workshops to investigate specific problems of privacy policy that CPPs could solve~\cite{zhang2025privcaptcha,reinhardt2021visual}. For this formative process, we engaged experienced designers and privacy \& security researchers to collectively generate potential interaction designs. The sessions were structured to progressively converge on a single, promising candidate from the initial proposals, based on evaluation of their effectiveness and feasibility.

\subsection{Study Setup}

\paragraph{Recruitment and Participants}

% We recruited 10 experienced participants (4 males, 6 females, with a mean age of 26.7, SD=5.0) through a hybrid method, combining personal contact, sending e-mails to authors of privacy policies papers in recent three years, and a snowball sampling~\cite{goodman1961snowball}. We focused on recruiting researchers of privacy and security, researchers of human computer interaction, and designers. Of the 10 participants, 6 identified them as privacy and security researchers, 6 identified them as human computer interaction researchers, and 1 self-identified them as designer. Participants on average published 16.8 (SD=30.5) top-tier conferences in 5 years, and have an averaged experience of 4.8 (SD=3.1) years. They self-rated their familiarity with user center design with a mean score of 6.9 (SD=1.4) on a ten-point scale. The study got approved by our Institutional Review Board (IRB) and each participant was compensated \$50 for the 2-hour long session.

We recruited 10 experienced participants (4 males, 6 females; $M_{age}=26.7$, $SD=5.0$) via personal contacts, e-mails to recent privacy-policy authors, and snowball sampling~\cite{goodman1961snowball}. Six self-identified as privacy/security researchers, six as HCI researchers, and one as designer. On average, they had 4.8 years of experience ($SD=3.1$), 16.8 publications ($SD=30.5$) in top-tier venues over 5 years, and rated their familiarity with user-centered design at 6.9/10 ($SD=1.4$). The study got approved by our Institutional Review Board (IRB) and each participant was compensated \$50 for the 2-hour long session.

\paragraph{Rationale and Study Design}
% Rather than a single large focus group, we deliberately ran three workshops structured as mini-focus groups (3--4 participants each) to reduce social pressure, mitigate dominance effects, and increase per-participant airtime~\cite{cameron2005facilitation}. Multiple small sessions also enabled us to cross-validate themes across sessions.

Rather than one large focus group, we ran three workshops structured as mini-focus groups (3–4 participants) to reduce social pressure, mitigate dominance, and increase per-participant airtime~\cite{cameron2005facilitation}, while enabling cross-session validation.

All workshops followed the same moderator guide and materials for consistency. Before workshops, participants received a concise CPP primer, a figure from Pan et al.~\cite{pan2024new} contrasting CPP with runtime- and install-time notices, and an overview of the reflective-thinking framework with stage-specific prompts. We also provided two example CPP designs as walkthroughs of the reflective-thinking stages (\textit{experience} → \textit{reflection} → \textit{action}), explicitly framed as illustrative to avoid anchoring. As a pre-task, each participant prepared two CPP designs for mobile-app scenarios.

In the workshops, participants first shared and iterated on their designs through group discussion, focusing on interaction flows, problems addressed, and underlying design theories. They then discussed and prioritized the privacy-policy problems most suitable for CPP to address and shortlisted the most promising design(s) based on feasibility and effectiveness. Sessions concluded with open-ended feedback. All workshops were held online using Tencent Meeting\footnote{\url{https://meeting.tencent.com/}} or Zoom\footnote{\url{https://zoom.com/}}, with discussions and materials shared via a collaborative Figma\footnote{\url{http://figma.com/}} workspace. All sessions were audio- and video-recorded and later transcribed for analysis.

% We then transitioned to a convergence phase. The group clustered related designs, merged near-duplicates, and clarified boundaries between adjacent proposals. As an anchor for selection, participants first prioritized the privacy problem a CPP should address in the current scenario by offering immediate justifications for their choices. Using that prioritized problem, they then recorded individual judgments on each design's effectiveness and feasibility, discussed divergences to surface risks, constraints, and negative cases, and selected one most promising design per session alongside a within-session shortlist of two to three designs. For every selection, succinct rationales were documented.

% The workshop session concluded with an open-ended discussion to gather any additional feedback. The entire session was conducted online using Tencent Meeting\footnote{\url{https://meeting.tencent.com/}} or Zoom\footnote{\url{https://zoom.com/}}, with all discussions and materials shared via a collaborative Figma\footnote{\url{http://figma.com/}} workspace. The materials were audio- and video- recorded and later transcribed for analysis.

To synthesize across sessions while controlling bias, we subsequently ran a cross-session prioritization via an online survey platform. Shortlisted designs from all workshops were converted into anonymized idea cards, randomly ordered and blinded to session of origin. All participants independently and asynchronously ranked the cards for effectiveness and feasibility.

\paragraph{Analysis Process}

We adopted thematic analysis~\cite{braun2021thematic} on our transcribed text materials. The visual materials were first translated to text descriptions before analyzing, following prior guidance~\cite{bailey2008first}. One primary author first reviewed all the transcripts and constructed the initial codebook. Then the primary author and a secondary author discussed on the initial codebook and reached consensus. Then these two authors coded the transcripts and iteratively refined the codebook, with intermittent discussions to solve newly emerged disagreements. Given the inductive nature of this process and the documented drawbacks of calculating inter-rater reliability in this context~\cite{mcdonald2019reliability}, we ensured quality control through these intermittent discussions rather than by reporting inter-rater reliability scores. For the cross-session prioritization, we aggregated the ranking results using the Borda Count, a commonly adopted method for combining ranked preferences~\cite{mclean1995classics}.

\subsection{Results}

We present our findings in three parts: first, a summary of the key problems with current privacy policies that CPP could address, second, a design space for CPPs built on the reflective thinking framework, and third, an analysis of participant preferences for these designs and problem types. 

\subsubsection{Problems of Privacy Policies That CPP Could Solve}

We identified six problems (Pr1-6) of privacy policies that participants thought CPP could solve, where \textit{the disconnect between context and action}~\cite{pan2024new,windl2022automating}, and \textit{information overload and poor readability}~\cite{reinhardt2021visual,zhang2025privcaptcha} were readily documented. We additionally identified problems such as \textit{failure to communicate tangible risk} and \textit{erosion of user agency and motivation}, which were essential for designing CPPs, and motivated our reflective thinking-based framework (see Table~\ref{tab:main_ranking_summary}).

\textbf{Disconnect between context and action (Pr1):} Participants thought traditional policies are presented at the wrong time, such as at the initial sign-up, and in the wrong place, such as at a hidden legal page, making it impossible for users to connect the abstract text to their concrete actions within the app. This echoed previous paper's argument that CPPs could better correlate contextual risks with privacy policy segments~\cite{pan2024new}. P1 noted that policies are \textit{``always displayed on the sign-up page,''} but users rarely connect this to subsequent operations within the app's workflow. P3 echoed by explaining that policy text \textit{``lacks context''}.  

\textbf{Failure to communicate tangible risk (Pr2):} A key issue is that policies are written in abstract, legalistic language that describes procedures rather than communicating real-world consequences. P4 explicitly framed this as the need to move from ``procedure-based'' to ``risk-based'' notices. Instead of legal text, a user would want to know, \textit{``you may receive five more scam calls per month on average''} by providing a phone number. P5 stated the policy needs to explain the ``worst-case situation'', such as the risk of being stalked if the location information was shared. This motivated our reflective thinking-based design, and also echoed the trends~\cite{chen2025clear}.

\textbf{Erosion of user agency and motivation (Pr3):} Participants identified a critical need to combat users' feeling of powerlessness. When users feel they have no meaningful control, they lose the motivation to engage with privacy information at all, leading to ``privacy resignation''~\cite{draper2017privacy}. As P5 noted, a top priority is to make users feel \textit{``active in protecting their privacy.''} P7 explained that the lack of options beyond ``accept all data collection or stop using the app entirely'' causes users to \textit{``lose a lot of motivation to read the policy''}.

\textbf{Information overload and poor readability (Pr4):} Participants found policies too long, dense, and difficult to comprehend. This aligns with prior research, which found that users often ignore privacy policies due to these issues~\cite{reinhardt2021visual,zhang2025privcaptcha}, a problem that could be potentially mitigated by CPPs~\cite{pan2024new}. As P2 noted, privacy policies are \textit{``a very long text website''} that \textit{``most people will not have the patience to read''}. P9 hoped to design simple, visual cues to convey information ``at a glance'' without the need for ``lengthy text descriptions''. 

\textbf{Mismatch between policy and practice (Pr5):} Participants expressed concern that an app's actual data practices may not align with its stated policy. P4 mentioned the real-world ``rumors'' of apps selling personal information, a practice that is unlikely to be officially stated in policies. This suggests a need for designs that track app's behavior. 

\textbf{Outdated content for modern risks (Pr6):} Participants noted that static legal documents often fail to keep pace with modern technological advancements, particularly the rise of AI. P1 argued that many current policies are a ``product of the last decade.'' For example, a key piece of missing information is whether user data, such as photos, will be used for ``AI model training'', a modern risk users are largely unaware of.

\subsubsection{A Design Space For CPPs With Reflective Thinking Framework}

\begin{figure}[!htbp]  
    \centering
    \includegraphics[width=0.9\textwidth]{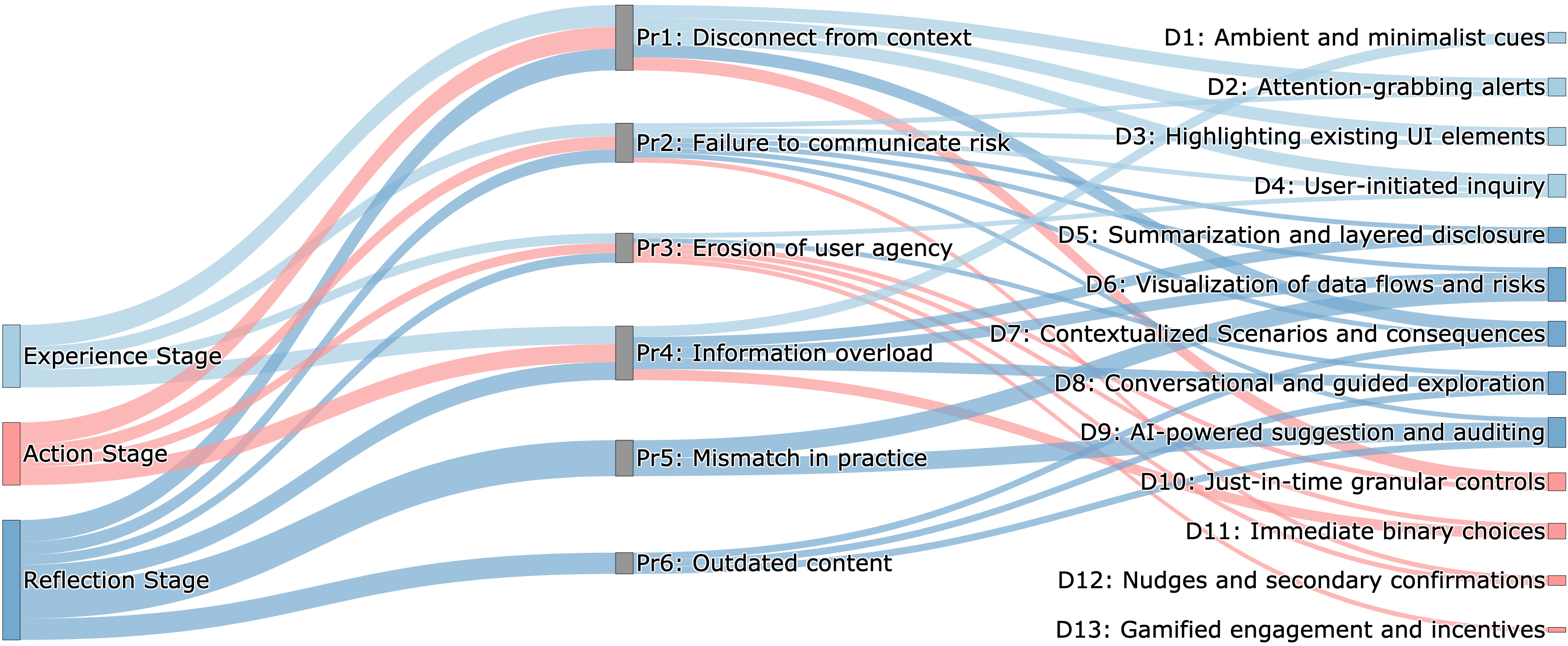}
    \caption{A sankey diagram showing the mapping relationship between the problems, the three reflective thinking stages, and each stage's designs, where different colors corresponded to different stages. The size of the problem's node indicated its rated importance in Section~\ref{sec:formative_study_problem}.}
    \label{fig:sankey}
\end{figure}

To systematically address the identified shortcomings of traditional privacy policies, we ground our design space in reflective thinking framework~\cite{schon1983reflective}. This framework allows us to structure CPP designs across three stages, experience, reflection and action, that collectively address all identified issues (see Figure~\ref{fig:sankey}). \textit{Experience stage} provide in-situ triggers about data practices, \textit{Reflection stage} translate abstract policies into understandable insights. \textit{Action stage} empower users with clear and timely controls.
% each directly corresponding to the problems we identified (see Figure~\ref{fig:designs}). The \textit{Experience stage} confronts the \textit{disconnect between context} by providing in-situ triggers about data practices. The subsequent \textit{Reflection stage} tackles the \textit{failure to communicate tangible risk} and \textit{information overload} by translating abstract policies into understandable insights. Finally, the \textit{Action stage} combats the \textit{erosion of user agency} by empowering users with clear and timely controls. The following sections detail the specific designs that operationalize each stage of this comprehensive framework.

\begin{figure}[!htbp]
    \subfloat[]{
        \includegraphics[width=0.25\textwidth]{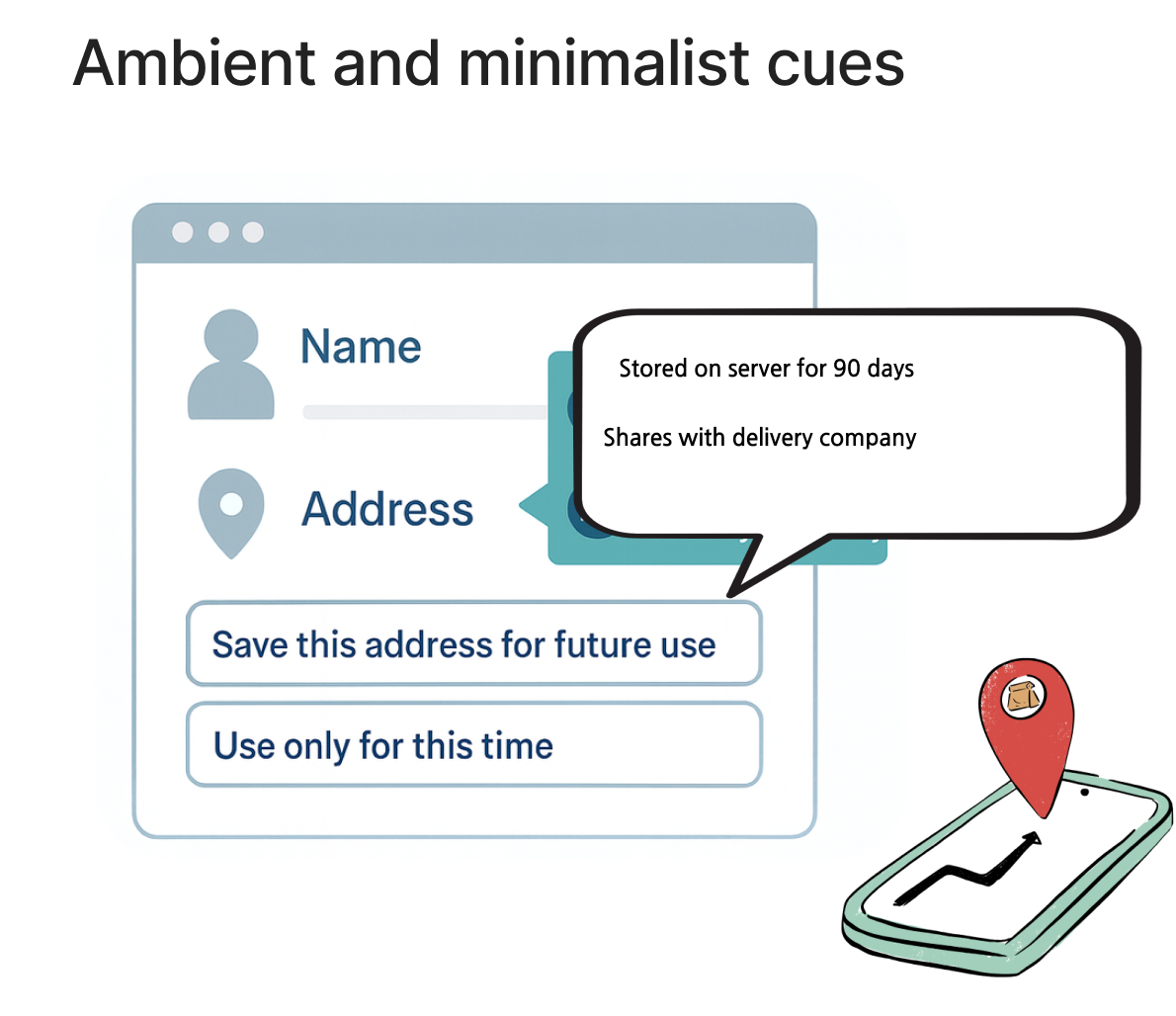}%
    }
    \subfloat[]{
        \includegraphics[width=0.25\textwidth]{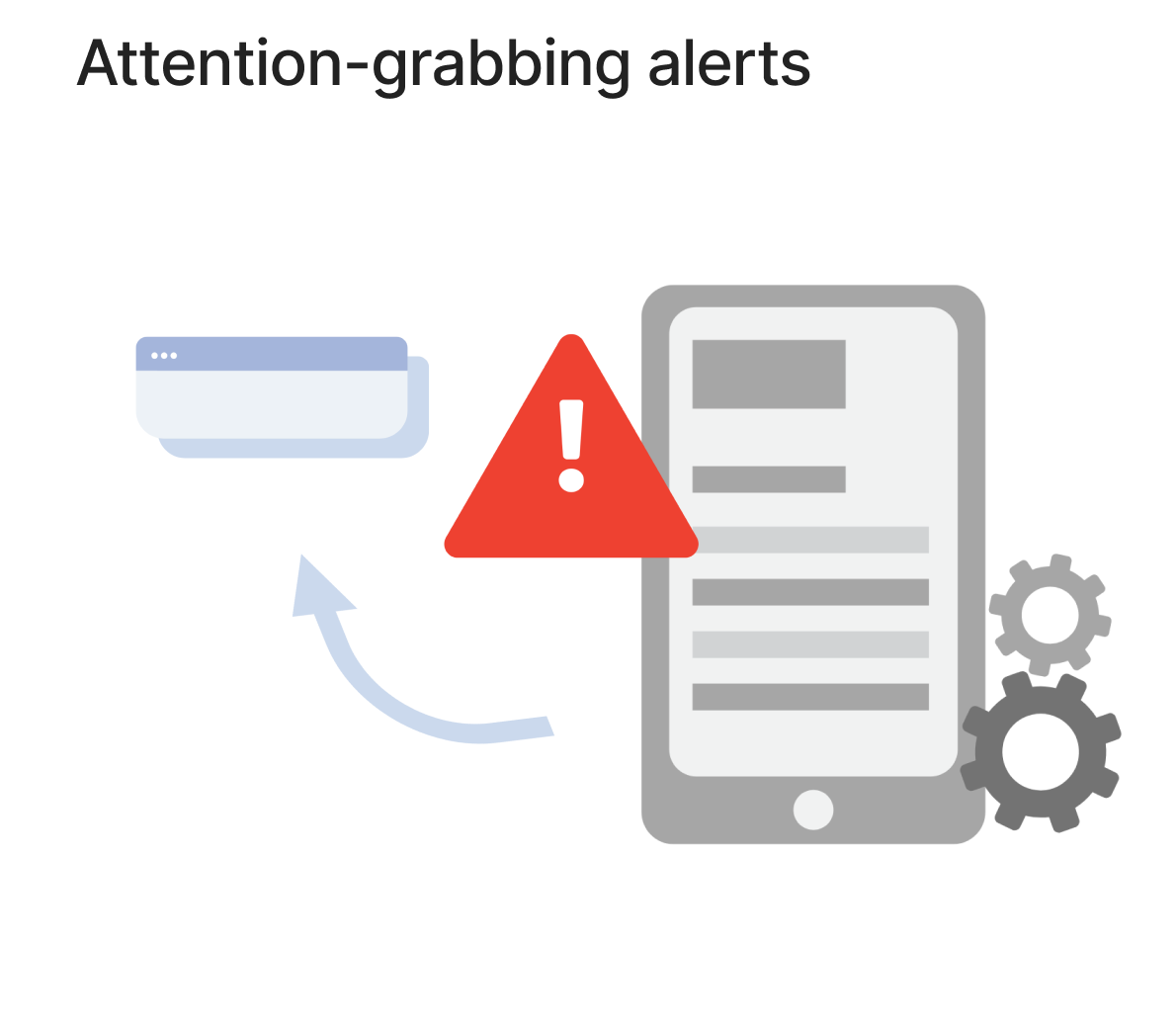}%
    }
    \subfloat[]{
        \includegraphics[width=0.25\textwidth]{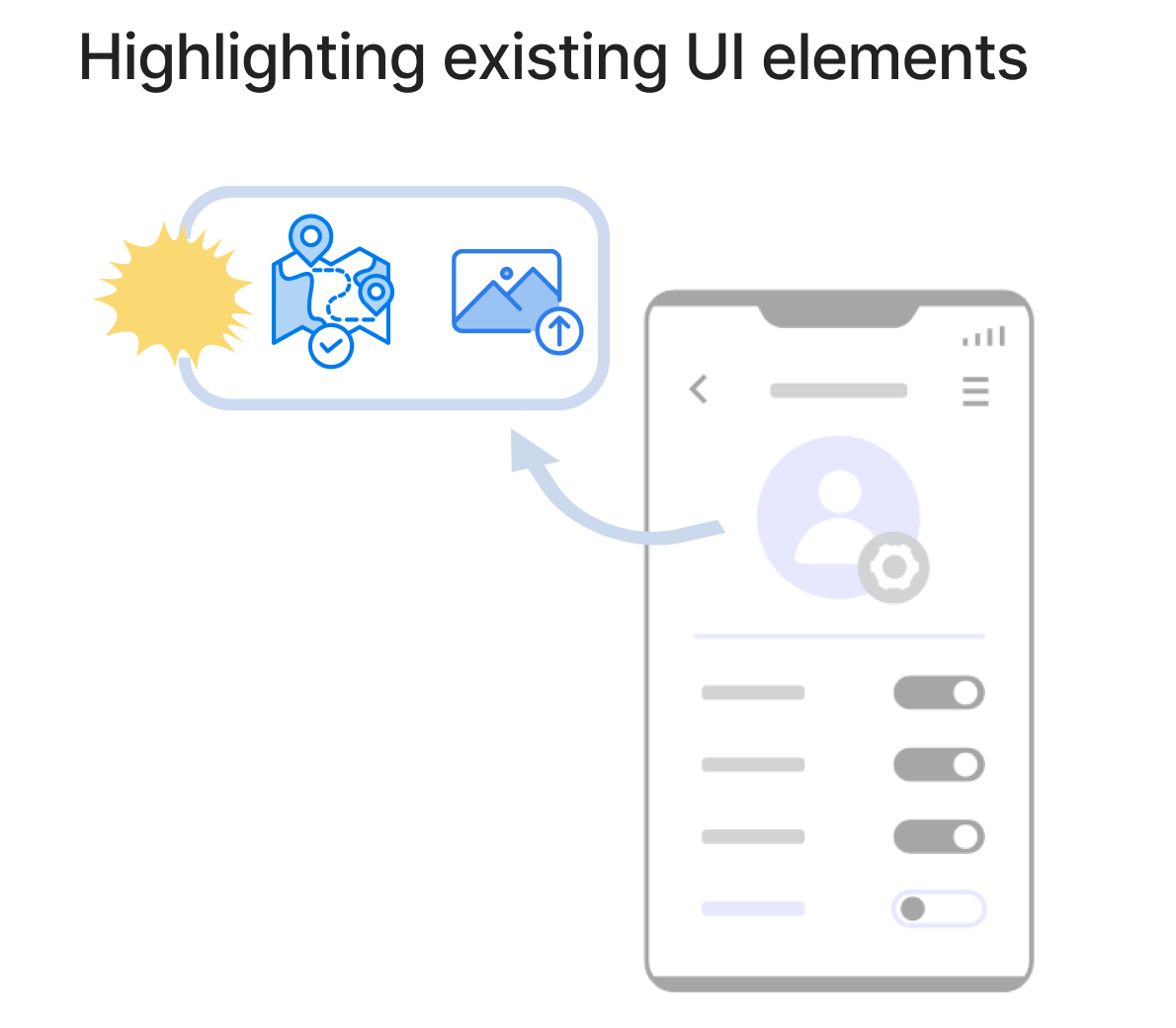}%
    }
    \subfloat[]{
        \includegraphics[width=0.25\textwidth]{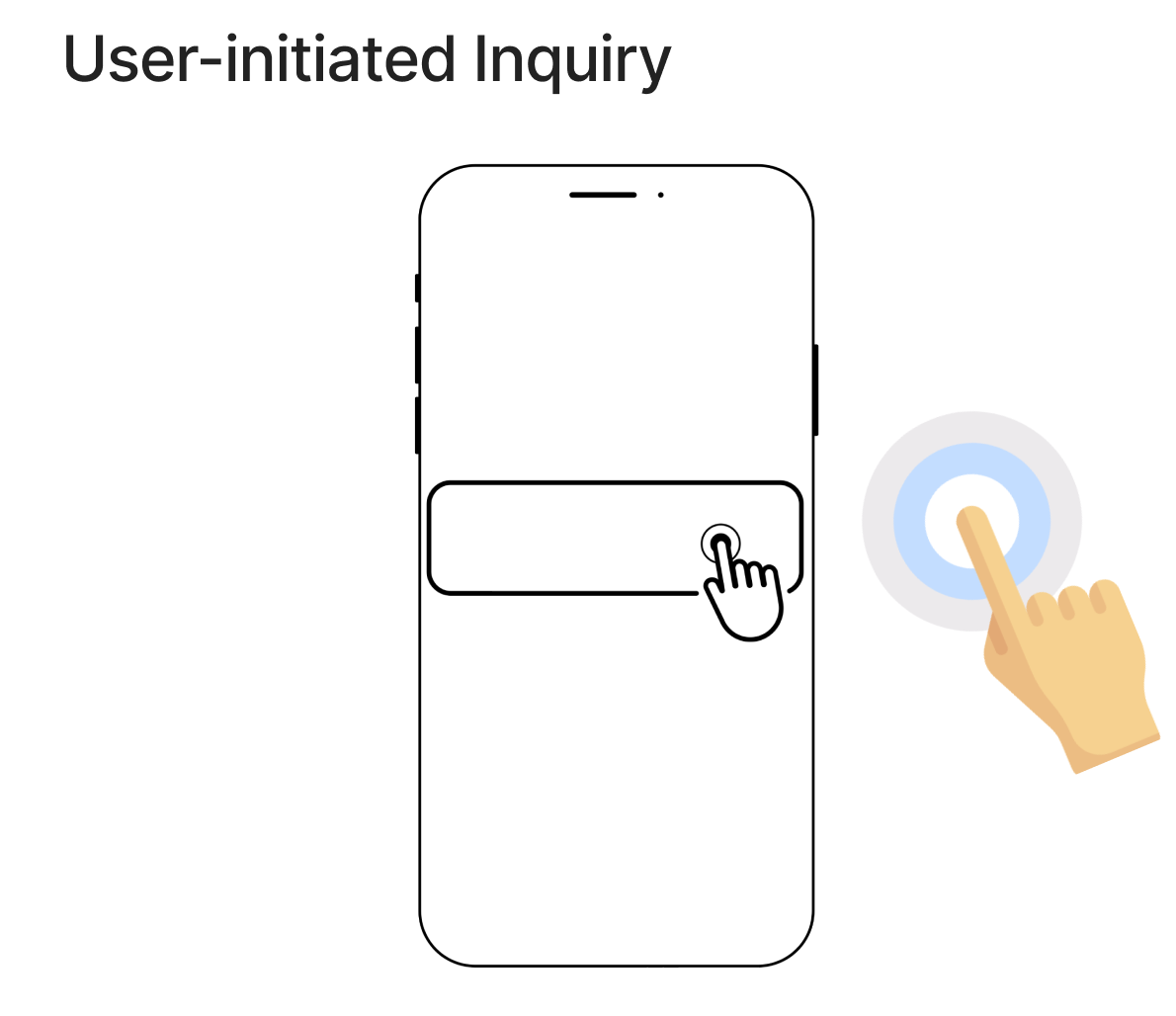}
    }

    \caption{Visualization of all designs in the experience stage, (a) Ambient and minimalist cues, (b) Attention-grabbing alerts, (c) Highlighting existing UI elements, (d) User-initiated inquiry.}
    % reflection (e,f,g,h,i) and action stage (j,k,l,m) respectively. (a) Ambient and minimalist cues, (b) Attention-grabbing alerts, (c) Highlighting existing UI elements, (d) User-initiated inquiry, (e) Summarization and layered disclosure, (f) Visualization of data flows and risks, (g) Contextualized scenarios and consequences, (h) AI-powered suggestion and auditing, (i) Conversational and guided exploration, (j) Just-in-time granular controls, (k) Nudges and secondary confirmations, (l) Immediate binary choices, and (m) Gamified engagement and incentives.}
    \label{fig:design_experience}
\end{figure}

\paragraph{Experience Stage}

This dimension directly confronts the problems such as \textit{disconnect between context and action} (Pr1) by providing timely, in-situ triggers about data practices, thereby mitigating \textit{information overload} (Pr4).

% This dimension focuses on how users are initially made aware of data practices within a specific context, serving as a trigger for the reflective process. 

\textbf{D1: Ambient and minimalist cues.} These designs use subtle, low-interruption indicators integrated into the user interface, \textit{mitigating information overload} (Pr4). They signal the presence of a privacy-relevant action without demanding immediate attention, prioritizing an unobtrusive user experience. For example, P3 proposed a numerical indicator next to a data icon to show how many times a data type is mentioned in a policy, while P10's design used a small, floating card next to a data entry field as a quiet hint. P9's design employed non-intrusive icons next to an address field during checkout to inform users about data handling.

\textbf{D2: Attention-grabbing alerts.} This approach uses prominent visual cues to actively capture user attention, particularly for high-stake data transactions, thereby \textit{highlighting tangible risk} (Pr2). It prioritizes immediate awareness over minimizing interruption. P5's design, for instance, used a red exclamation mark icon for non-essential or intrusive data collection. P1's design visualized risk levels with a color-coded system, such as green, orange, and red, to immediately alert users to the potential consequences of their selections.

\textbf{D3: Highlighting existing UI elements.} This design makes users aware of privacy implications by drawing attention to elements already present in the interface that are often overlooked (Pr1). P6's design highlighted on-screen personal information, like a child's name in a trip history, making the abstract concept of personal data tangible. P9's design used a flashing effect to highlight an app's native privacy settings when they became relevant, guiding the user toward an existing control.

\textbf{D4: User-initiated inquiry.} Unlike proactive notifications, this design empowers users to actively seek out privacy information in their own terms. It provides an on-demand tool for users with a specific privacy concern. For examples, P7's design featured a system-level floating button that users could drag onto any UI element to retrieve its relevant privacy information.

\paragraph{Reflection Stage}

This dimension focuses on translating abstract policy into understandable insights, tackling all problems, such as the \textit{failure to communicate tangible risk} (Pr2), \textit{information overload} (Pr4), \textit{mismatch between policy and practice} (Pr5) and the \textit{outdated content} (Pr6).

\begin{figure}[!htbp]

    \subfloat[]{
        \includegraphics[width=0.2\textwidth]{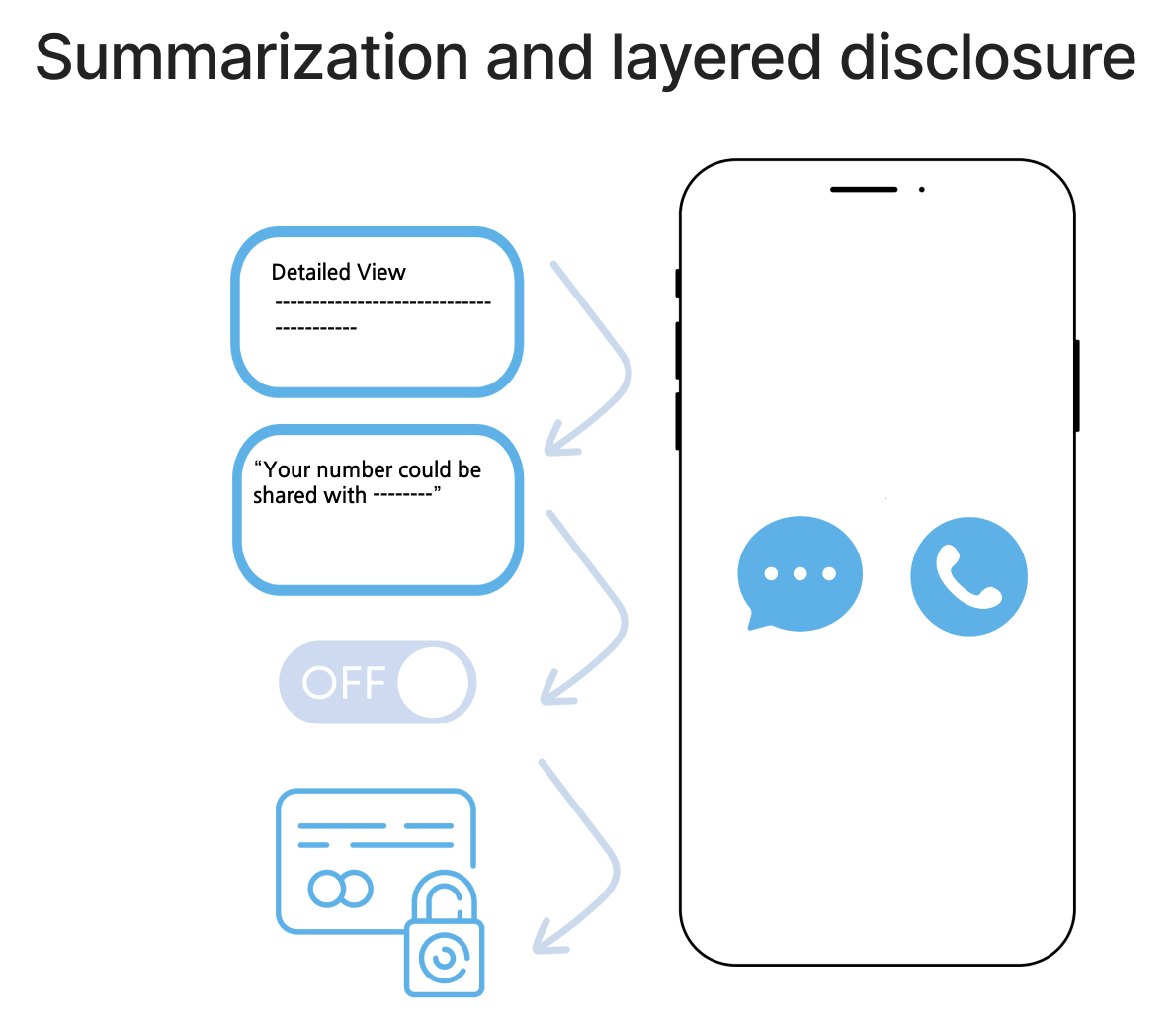}
    }
    \subfloat[]{
        \includegraphics[width=0.2\textwidth]{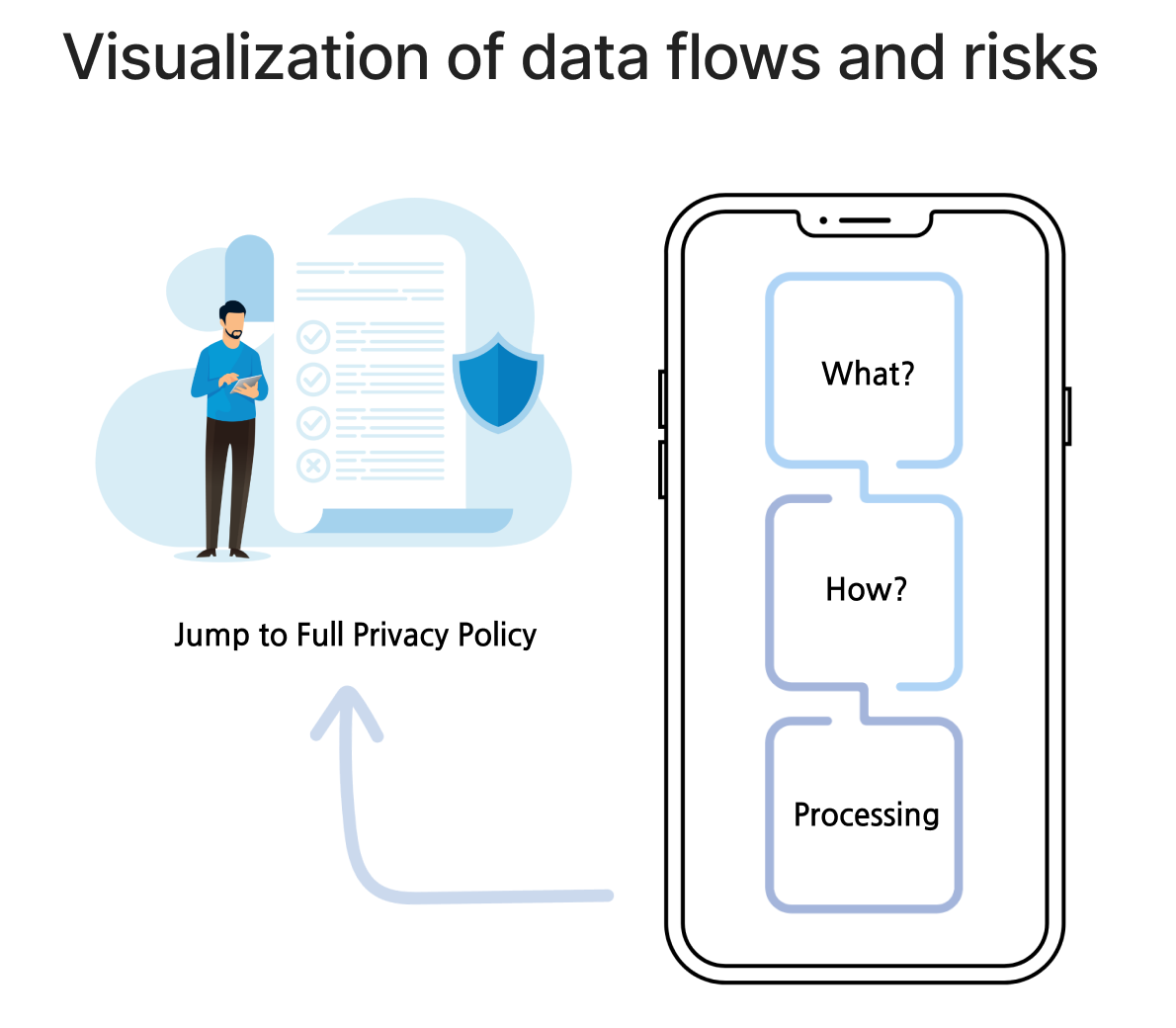}
    }
    \subfloat[]{
        \includegraphics[width=0.2\textwidth]{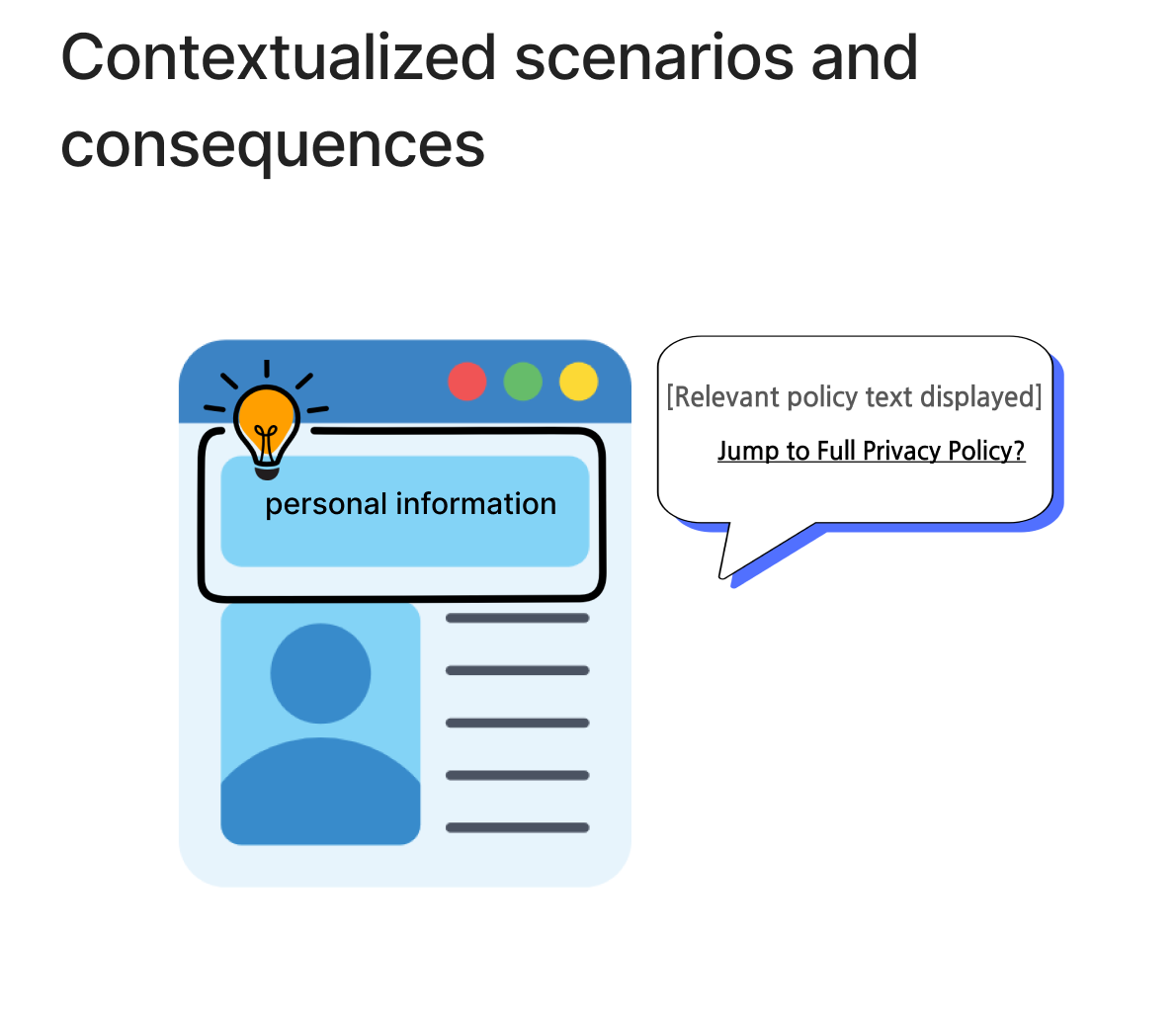}
    }
    \subfloat[]{
        \includegraphics[width=0.2\textwidth]{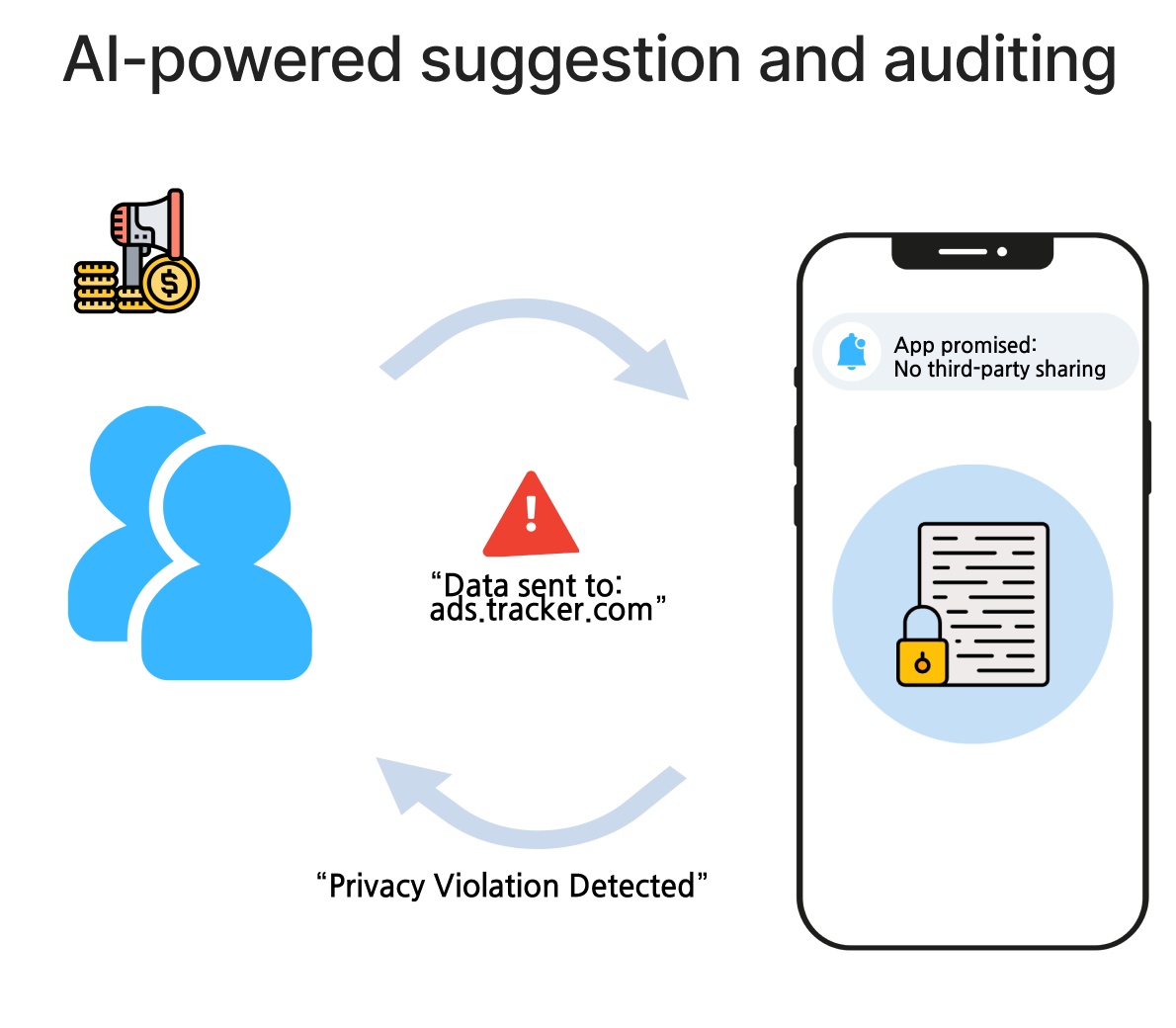}
    }
    \subfloat[]{
        \includegraphics[width=0.2\textwidth]{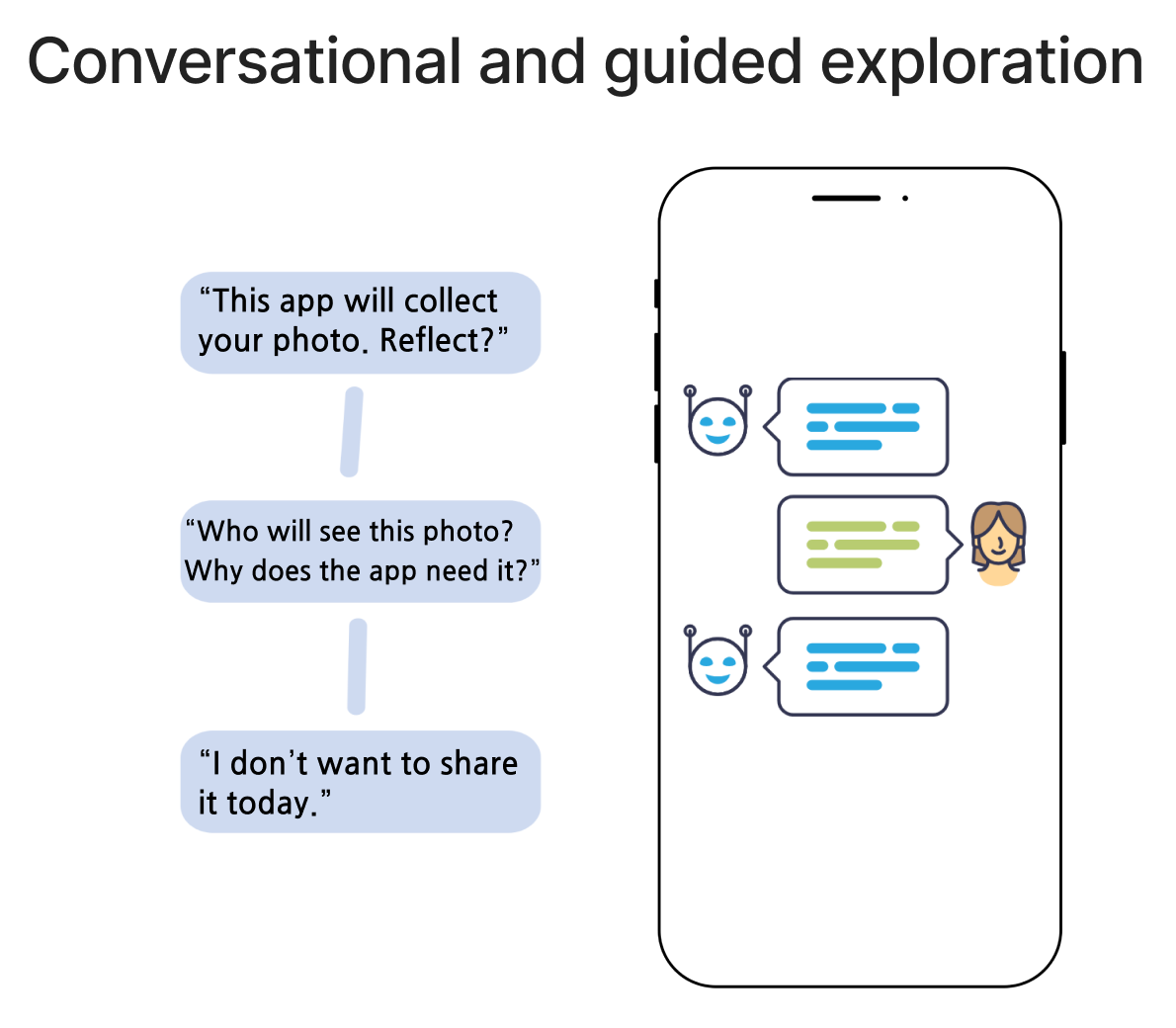}
    }

    \caption{Visualization of all designs in the reflection stage, (a) Summarization and layered disclosure, (b) Visualization of data flows and risks, (c) Contextualized scenarios and consequences, (d) AI-powered suggestion and auditing, (e) Conversational and guided exploration. }
    % reflection (e,f,g,h,i) and action stage (j,k,l,m) respectively. (a) Ambient and minimalist cues, (b) Attention-grabbing alerts, (c) Highlighting existing UI elements, (d) User-initiated inquiry, (e) Summarization and layered disclosure, (f) Visualization of data flows and risks, (g) Contextualized scenarios and consequences, (h) AI-powered suggestion and auditing, (i) Conversational and guided exploration, (j) Just-in-time granular controls, (k) Nudges and secondary confirmations, (l) Immediate binary choices, and (m) Gamified engagement and incentives.}
    \label{fig:design_reflection }
\end{figure}

% This dimension focuses on the content and format of privacy information presented after the initial trigger. The goal is to translate abstract privacy policy language to understandable insights that enable users to assess risks and consequences. 

\begin{table}[htbp]
    \centering
    \caption{The aggregated ranking scores (Borda Count) of different aspects.}
    \label{tab:main_ranking_summary}
    % 使用resizebox确保整个大表格不会超出页面宽度
    \subfloat[Designs of experience stage.]{
    \resizebox{0.47\textwidth}{!}{
    
        \label{tab:experience}
        \begin{tabular}{rlr}
            \toprule
            Agg. Rank & Item & Score \\
            \midrule
            1 & Highlight existing UI elements & 20 \\
            2 & Ambient and minimalist cues & 19 \\
            3 & Attention-grabbing alerts & 11 \\
            4 & User-initiated inquiry & 10 \\
            \bottomrule
        \end{tabular}
    }}%  <-- 这个 '%' 很重要，防止产生多余空格
    \hfill         %  <-- 让两个表格左右对齐
    \subfloat[Designs of reflection stage.]{
    \resizebox{0.47\textwidth}{!}{
        
        \label{tab:reflection}
        \begin{tabular}{rlr}
            \toprule
            Agg. Rank & Item & Score \\
            \midrule
            1 & Contextualized scenarios and consequences & 32 \\
            2 & Summarization and layered disclosure & 24 \\
            3 & Visualization of data flows and risks & 23 \\
            4 & AI-powered suggestion and auditing & 11 \\
            5 & Conversational and guided exploration & 10 \\
            \bottomrule
        \end{tabular}
    }}%  <-- 这个 '%' 很重要

    % --- 第二行子表格 ---
    \subfloat[Designs of action stage.]{
    \resizebox{0.47\textwidth}{!}{
        \centering
        \label{tab:action}
        \begin{tabular}{rlr}
            \toprule
            Agg. Rank & Item & Score \\
            \midrule
            1 & Just-in-Time granular controls & 23 \\
            2 & Immediate binary choices & 18 \\
            3 & Nudges and secondary confirmations & 14 \\
            4 & Gamified engagement and incentives & 5 \\
            \bottomrule
        \end{tabular}
    }}%  <-- 这个 '%' 很重要
    \hfill         %  <-- 让两个表格左右对齐
    \subfloat[Problems.]{
    \resizebox{0.48\textwidth}{!}{
        \centering
        \label{tab:problem}
        \begin{tabular}{rlr}
            \toprule
            Agg. Rank & Item & Score \\
            \midrule
            1 & Disconnect between context and action & 40 \\
            2 & Information overload and poor readability & 33 \\
            3 & Failure to communicate tangible risk & 24 \\
            4 & Mismatch between policy and practice & 22 \\
            5 & Erosion of user agency and motivation & 18 \\
            6 & Outdated content for modern risks & 13 \\
            \bottomrule
        \end{tabular}
        }
    }%  <-- 这个 '%' 很重要
\end{table}

\textbf{D5: Summarization and layered disclosure.} Directly addressing \textit{information overload and poor readability} (Pr4), this approach structures information into digestible summaries. For example, P3 categorized policy snippets by their original section, while P2 proposed breaking policies into collapsible, topic-based sections. P10's design featured a multi-step disclosure, starting with a simple prompt and expanding into a detailed summary and control option. P1's design deconstructed abstract permission requests into clear, bulleted points.
% This approach structures complex privacy information by breaking it into digestible summaries and presenting it in progressive layers. This allows users to control the depth of information they consume. 

\textbf{D6: Visualization of data flows and risks.} This design leverages graphical elements to communicate complex data practices intuitively, thereby reducing \textit{information overload} (Pr4) and making \textit{risks more tangible} (Pr2). For example, P9's design used visualizations like maps to show how location data could reveal travel patterns and flowcharts to illustrate the movement of information. P8 suggested using visual data flowcharts to show where information goes, and P1's design employed color-coding as a direct visualization of privacy risk levels. 
%in an intuitive and immediately understandable format

\textbf{D7: Contextualized scenarios and consequences.} This design moves beyond procedural descriptions to focus on risk-based narratives. It uses relatable scenarios and worst-case examples to make the consequences of data sharing more tangible and impactful, thereby addressing the \textit{failures to communicate tangible risks} (Pr2) and can be updated to address \textit{outdated content for modern risks} (Pr6). P4's design centered on ``risk-based notices'' that answered user-driven questions, such as why friends from one app appear on another. P5's design aimed to inform users about ``worst-case scenarios'' like bring targeted by stalkers, and P6 proposed generating story-like warnings to provide an emotional understanding of potential harm. 

\textbf{D8: Conversational and guided exploration.} This design uses interactive, dialogue-based interfaces, such as chatbots or assistants, to engage users in a natural exploration of privacy issues, overcoming \textit{information overload} (Pr4). P10's chatbot proactively guided user queries with pre-generated questions. P2 envisioned an ``assistant-like'' interaction to explain how settings are used for ad targeting and guide users to relevant controls.

\textbf{D9: AI-powered suggestion and auditing.} This design uses computational methods to analyze context, policy and app behavior to provide actionable recommendations or verify the trustworthiness of an app's claims, thereby addressing the {mismatch between policy and practice} (Pr5) and \textit{outdated content} (Pr6). P3 proposed a model that would analyze a user's task and suggest whether a permission is necessary, such as ``location is not needed for browsing''. P7's design was a proactive system that monitored an app's network traffic and compared it against its stated policy. 

\paragraph{Action Stage}

% This dimension concerns the mechanisms provided for users to act upon their reflections. It focuses on how users can exercise control over their data in a timely and effective manner.
This stage directly combats the problems such as \textit{erosion of user agency} (Pr3) and motivation by providing mechanisms for user to exercise meaningful control. 

\begin{figure}[!htbp]

    \subfloat[]{
        \includegraphics[width=0.25\textwidth]{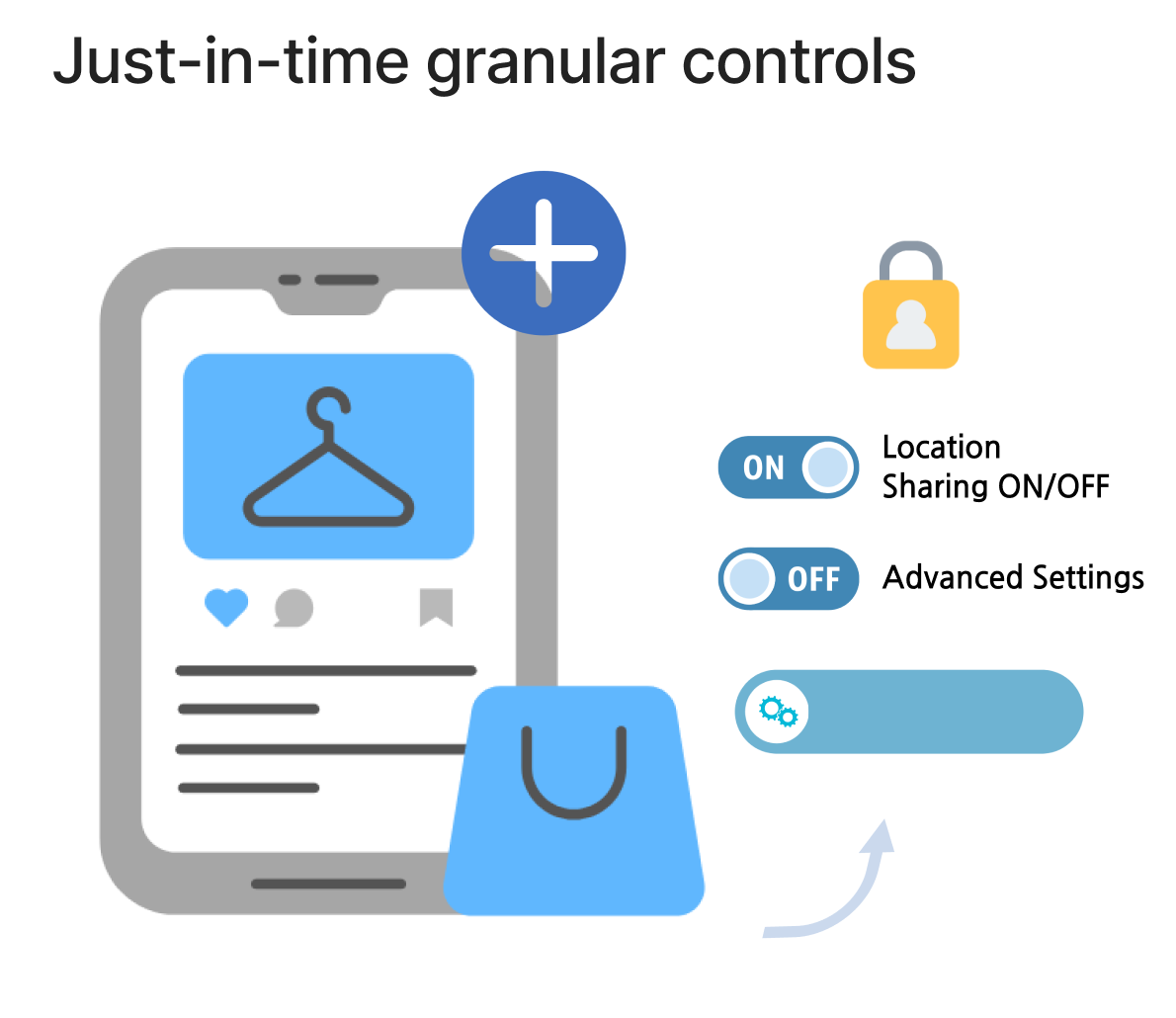}
    }
    \subfloat[]{
        \includegraphics[width=0.25\textwidth]{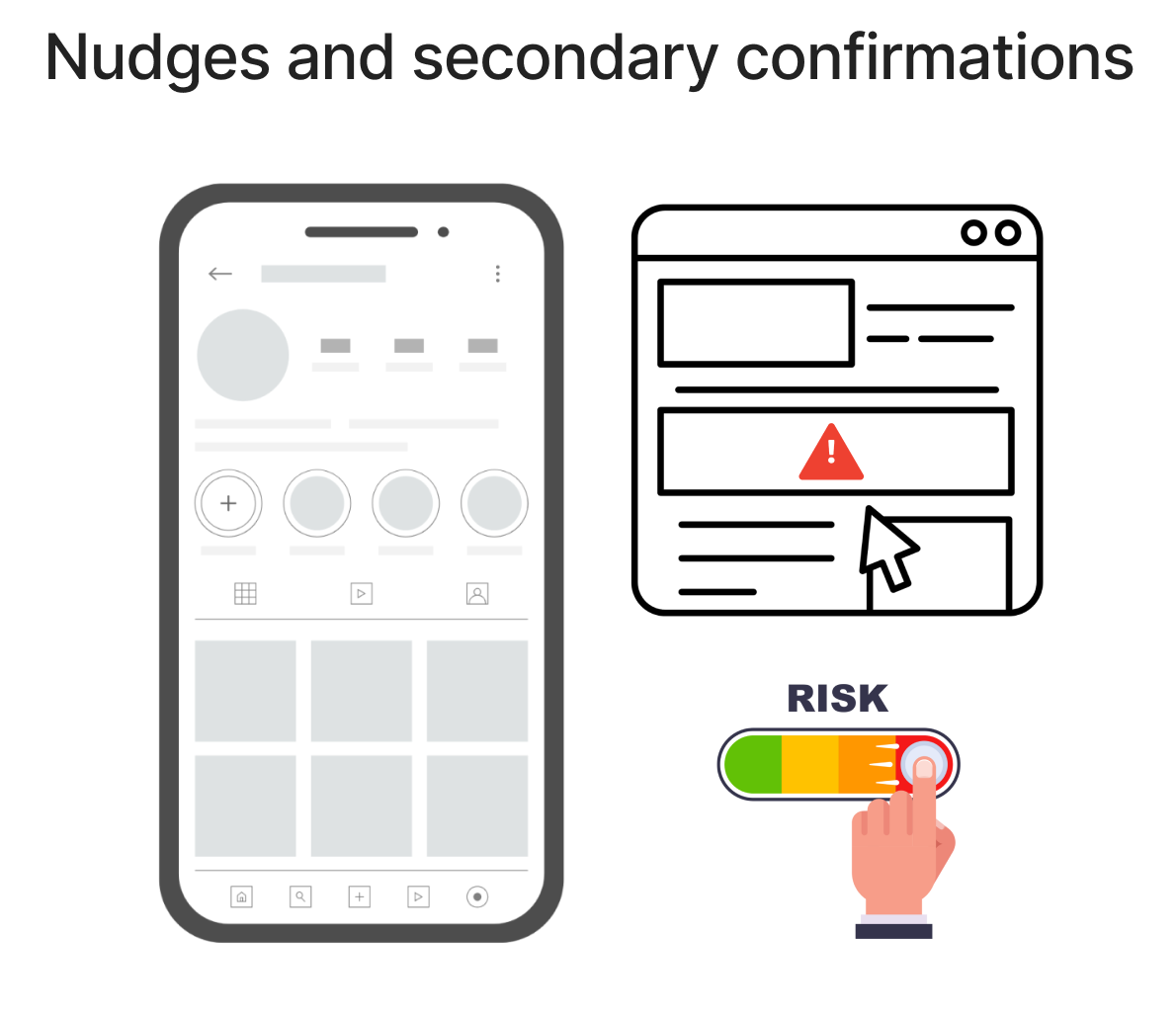}
    }
    \subfloat[]{
        \includegraphics[width=0.25\textwidth]{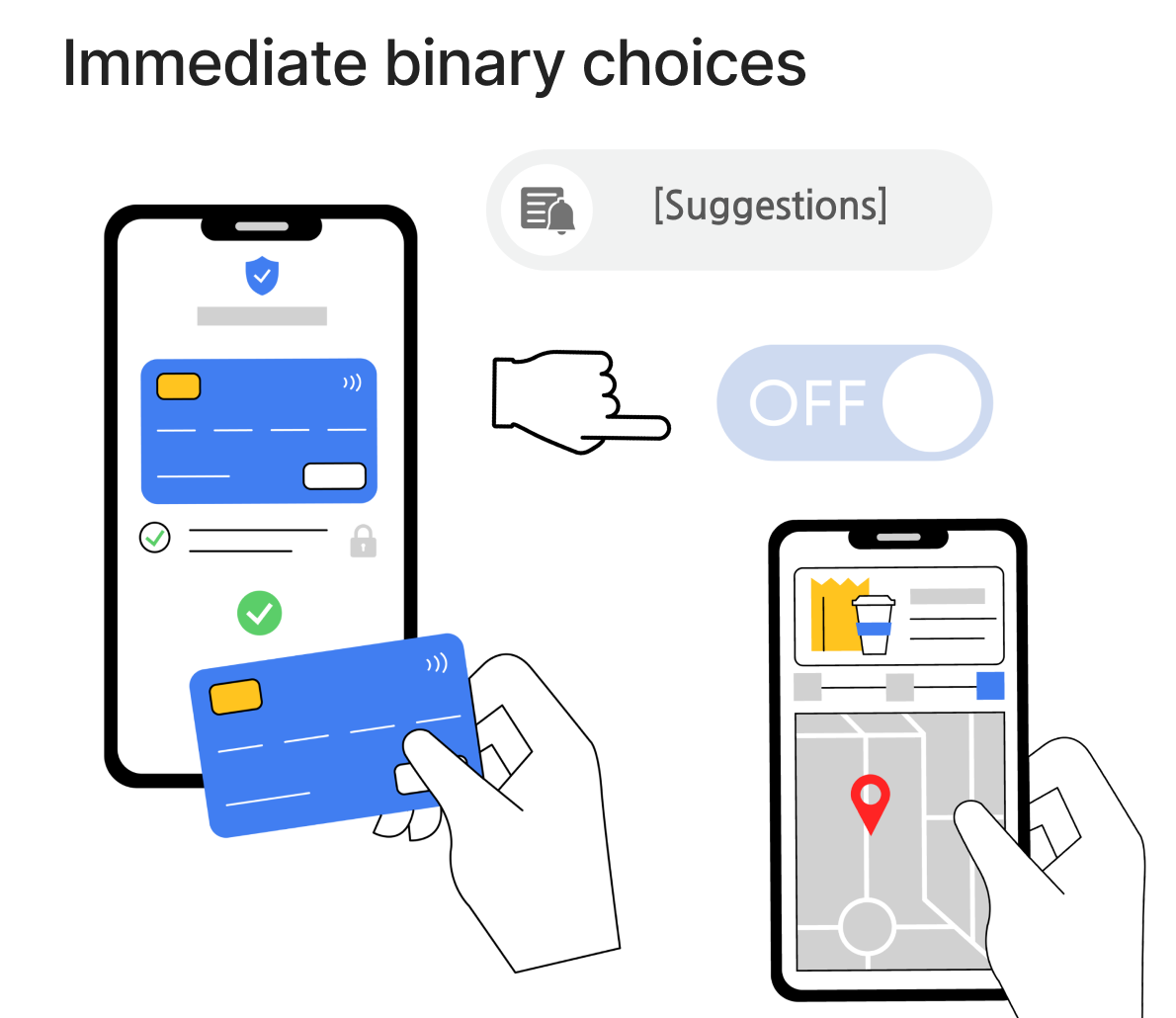}
    }
    \subfloat[]{
        \includegraphics[width=0.25\textwidth]{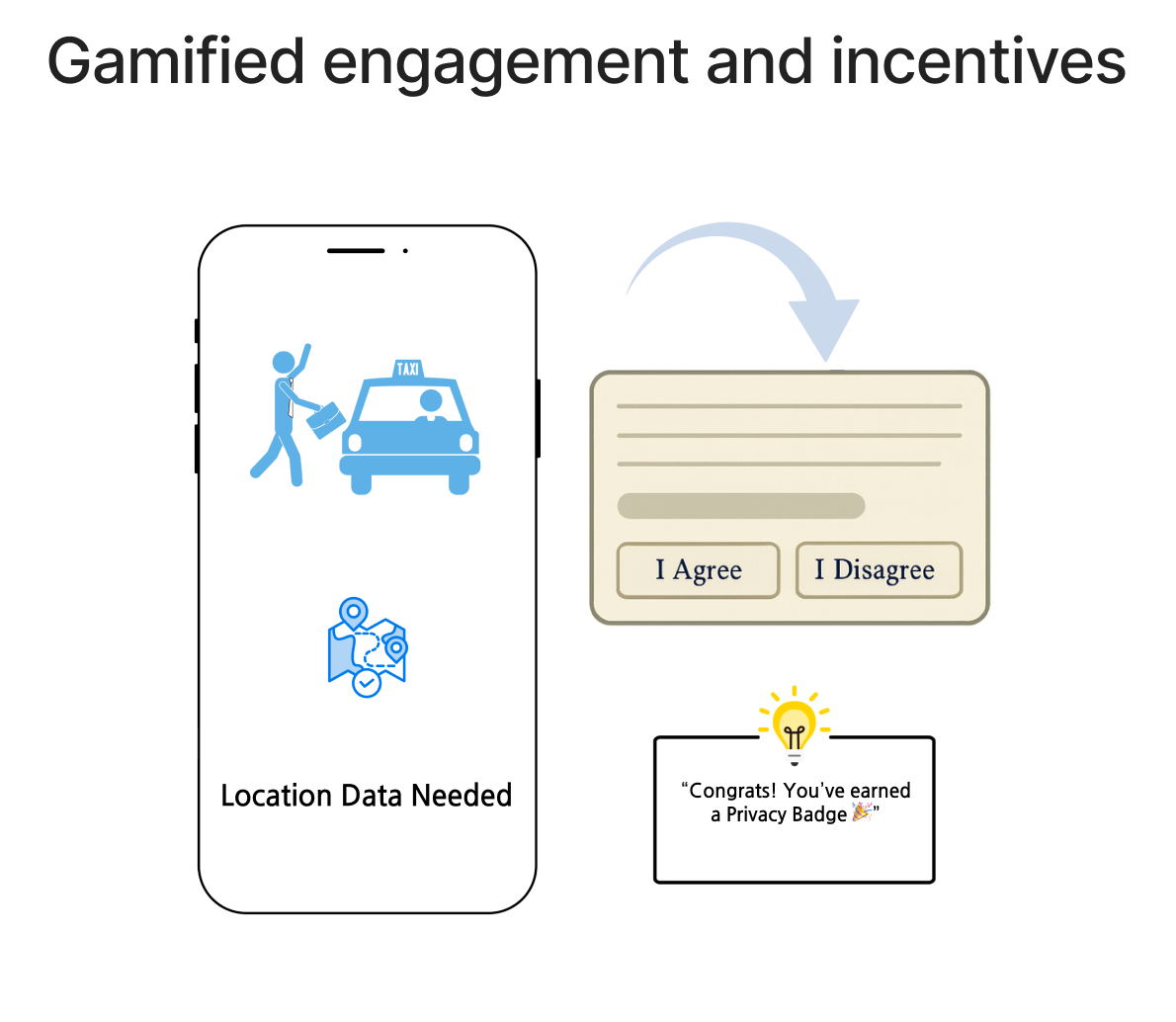}
    }
    \caption{Visualization of all designs in the action stage, (a) Just-in-time granular controls, (b) Nudges and secondary confirmations, (c) Immediate binary choices, and (d) Gamified engagement and incentives. }
    % reflection (e,f,g,h,i) and action stage (j,k,l,m) respectively. (a) Ambient and minimalist cues, (b) Attention-grabbing alerts, (c) Highlighting existing UI elements, (d) User-initiated inquiry, (e) Summarization and layered disclosure, (f) Visualization of data flows and risks, (g) Contextualized scenarios and consequences, (h) AI-powered suggestion and auditing, (i) Conversational and guided exploration, (j) Just-in-time granular controls, (k) Nudges and secondary confirmations, (l) Immediate binary choices, and (m) Gamified engagement and incentives.}
    \label{fig:design_action}
\end{figure}

\textbf{D10: Just-in-time granular controls.} These designs provide users with specific, fine-grained control options directly relevant to their current context, moving beyond simple on/off permissions. P9's design, for example, offered tiered control options for photo uploads, such as remove metadata before uploading, while P2 integrated a privacy settings button directly next to an address field for immediate, contextual adjustments. P1 advocated for granular settings within the permission prompt itself. 

\textbf{D11: Immediate binary choices.} This approach provides straightforward, often binary controls that are easy to understand and execute, facilitating quick decisions. P3's design included a prominent ``Turn Off'' button to act on a suggestion to disable a permission. Similarly, P10's design concluded its layered disclosure with an option to toggle a permission on or off, and P8's emphasized clear ``I agree'' or ``I disagree'' buttons within summary cards.

\textbf{D12: Nudges and secondary confirmations.} This design is intended to prevent user error or impulsive decisions by prompting users to reconsider an action, especially one with high privacy risk. P1's design implemented a secondary confirmation pop-up that appeared after a user attempted to post with a high-risk location-sharing setting, forcing a moment of re-evaluation.

\textbf{D13: Gamified engagement and incentives.} This novel approach uses game-like mechanics and external incentives to motivate users to engage with privacy information and controls, addressing user apathy. For example, P8's design suggests using elements like a progress bar for reading privacy cards or rewarding users with badges or discount coupons for completing a privacy-related task or quiz.

\subsubsection{Prioritization of Problems and Designs}\label{sec:formative_study_problem}
% Table~\ref{tab:main_ranking_summary} summarizes the aggregated scores from participant rankings, revealing distinct preferences. 
Overall, participants' design preferences strongly aligned with the problems they deemed most important (see Table~\ref{tab:main_ranking_summary}). Participants identified the \textit{disconnect between context and action} (score: 40) as the most critical issue, which aligns with existing literature~\cite{pan2024new} and our design framework. This was followed by \textit{information overload and poor readability} (score: 33)~\cite{reinhardt2021visual,zhang2025privcaptcha} and \textit{failure to communicate tangible risk} (score: 24). The fair level of consensus (Kendall's W = 0.31) underscores a shared perspective that the lack of context is the primary failing of current privacy policies.

Across all three stages, designs that provided context and control were consistently ranked highest. In the experience stage, participants slightly preferred \textit{highlighting existing UI elements} (score: 20) over \textit{ambient and minimalist cues} (score: 19). Both of these subtle, context-aware approaches were rated significantly higher than \textit{attention-grabbing alerts} (score: 11) or user-driven ones like \textit{user-initiated inquiry} (score: 10) (Kendall's W = 0.32, indicating fair level of agreement). 

In the reflection stage, \textit{contextualized scenarios and consequences} (score: 32) was strongly favored. This highlights a clear desire for information that directly translates abstract data practices into tangible, real-world outcomes. The next preferred designs, \textit{summarization and layered disclosure} (score: 24) and \textit{visualization of data flows and risks} (score: 23), were ranked closely together, demonstrating the value of making complex information digestible (Kendall's W = 0.35, indicating fair level of agreement).

In the action stage, \textit{just-in-time granular controls} (score: 23) was the most preferred design, emphasizing the value of providing specific, actionable choices at the exact moment a decision is needed. It was ranked notably higher than simpler designs like \textit{immediate binary choices} (score: 18) or \textit{nudges and secondary confirmations} (score: 14)(Kendall's W = 0.35, indicating fair level of agreement). 

\section{Design and Implementation of \proj{}}

This section proposed the design and implementation of the most prominent design in the workshops, instantiated as \proj{}, a prototype to demonstrate different apps' privacy policy contextually and in a reflective manner. We detailed its interface, interaction design and implementation details, and finally concluded with the technical evaluation.

\subsection{Design Goals}

Based on the formative study and prior work~\cite{zhang2025privcaptcha,reinhardt2021visual}, we established the following design goals:

\textbf{G1. Condense lengthy privacy policy and improve understanding (corresponding to Pr4).} CPPs aims to condense the lengthy privacy policy for users to understand the privacy policies easily. This could simplify the understanding barrier of privacy policies~\cite{reinhardt2021visual}. 
% , and was also highlighted in the prior work
% By integrating the condensed data practices of the specific categories into the data collection point on the screen, CPP could help users make sense of the privacy policy text. 

\textbf{G2. Facilitate meaningful and reflective engagement (corresponding to Pr1, Pr2, Pr3, Pr5, Pr6).} It is often argued that the privacy policy visualization only helps users to condense the information, but not to attract them, leading to user neglect. Therefore, CPPs should facilitate meaningful and reflective engagement.
% many users still would skip these visualizations without truly notice and understand the policies. To achieve this aim, CPPs could use methods such as mandatory interaction~\cite{zhang2025privcaptcha} or interactive demonstration~\cite{stellmacher2022escaping}.

\textbf{G3. Balance task completion and privacy education.} Although CPPs have the benefit of demonstrating privacy practices to users, users may have primary tasks to complete, which may conflict with the aim of privacy education~\cite{stellmacher2022escaping}. CPP should therefore balance task completion and privacy education.
% To achieve this aim, we argue that the CPP should balance task completion and privacy education. The contextual hint should allow users to directly complete task, and also attract their attention, and provide them chances to reflect on the privacy problems.

\subsection{Interaction Design}\label{sec:interaction_design}

The design of \proj{} is grounded in the three-stage reflective thinking framework~\cite{seaman2008experience,driscoll1994reflective,rolfe2001critical}, chosen for its balance of simplicity and effectiveness. We structured the users' interaction into a cycle of \textit{Experience}, \textit{Reflection} and \textit{Action}. 

%The user begins with the \textit{Experience Stage}. We designed a peripheral sidebar to be both a subtle trigger and a non-intrusive element, aligning with our goal of balancing task completion and privacy education (G3). This sidebar can be collapsed to minimize intrusion (Figure~\ref{}), and when new privacy risks are detected, it pops up with color-coded icons to indicate the level of data sensitivity. The color-coding formula is as follows: XXX. This visual cue prompts the user to consider the privacy implications without demanding immediate attention (G2,G3). %

The user begins with the \textit{Experience Stage}. We designed a floating window positioned at the screen edge as the default interface (D1), which surfaces detected privacy risks directly with color-coded icons indicating their sensitivity levels (Figure~\ref{fig:teaser_2}). The color-coding formula followed sensitivity results from previous practices~\cite{zhang2024ghost}. In this scheme, we mapped risk levels to a traffic-light style schema to provide intuitive feedback for general users: red indicates high risk, orange medium risk, and green low risk. This makes risks readily visible in a condensed form, helping users quickly grasp what type of data is involved and how sensitive it is (G1), while remaining lightweight and non-intrusive to ongoing tasks (D1, G3). For users who prefer minimal intrusion, the window can be collapsed via a toggle at the bottom into a small trigger (Figure~\ref{fig:teaser_1}). In the collapsed state, new risks initiates a  visual blinking animation that highlights the trigger to draw peripheral attention. These mechanisms together ensure that users receive timely cues to consider privacy implications without demanding immediate attention (G2,G3).

Upon a short click on a risk icon, the system transitions to the \textit{Reflection Stage} (G3). This triggers on-screen notifications, similar to system pop-ups~\cite{sanmorino2018design}, that provide just-in-place signaling with bounding boxes around relevant UI elements~\cite{pan2024new} (D5). The notifications offer a two-part, sequential risk-based disclosure: the first notification presents condensed data practice descriptions that highlight key aspects, including data collection, sharing and disclosure, as prior guidance~\cite{zhang2025privcaptcha,reinhardt2021visual} and practice~\cite{pan2024new,windl2022automating} indicated these are important aspects. The second notification provides reflective prompts that encourage users to consider the potential risks (D7, see Figure~\ref{fig:teaser_3}). These risks are generated and displayed contextually and in a hypothesized manner, similar to the sandbox approaches warning users of the risks~\cite{chen2024empathy}. These notifications are crafted from the app's privacy policy (see Section~\ref{sec:algorithm_design}), transforming abstract legal text into contextualized scenarios (G1,G2). We empirically set that, the initial notifications disappear in three seconds, but the final notification remain for five seconds, similar to the frequency of refreshment in prior design~\cite{zhang2025privcaptcha}, which offered the user a choice to engage in deep reflection. To support this deep reflection, a click on the final notification displays the original privacy policy in a system card format, allowing users to ``dig deeper'' into the text (see Figure~\ref{fig:teaser_4}). This feature is crucial for legal compliance and for users who seek a detailed understanding after an initial prompt. 

Within \proj{}, the \textit{Action Stage} is initiated via a long press on the risk icon (see Figure~\ref{fig:teaser_5}). This lightweight action allows users to permanently mute the corresponding risk triggers, reducing notification fatigue while preserving flexibility (D10). While our current prototype does not directly implement real granular just-in-time permission controls, which we further discuss in Section~\ref{sec:discussion-control}, it encourages users to adjust their privacy preferences by pointing them toward the relevant settings in the host application, empowering user control (D12). In this way, the \textit{Action Stage} bridges reflection and adjustment without interrupting the task flow (G3). By letting users selectively filter which risks remain visible, it helps reduce information overload and condense lengthy privacy texts into manageable cues (G1). By requiring a deliberate choice of whether to dismiss or continue monitoring a risk, it sustains reflective engagement (G2).

\subsection{Algorithm Design}\label{sec:algorithm_design}

\begin{figure}[!htbp]
    \includegraphics[width=0.7\textwidth]{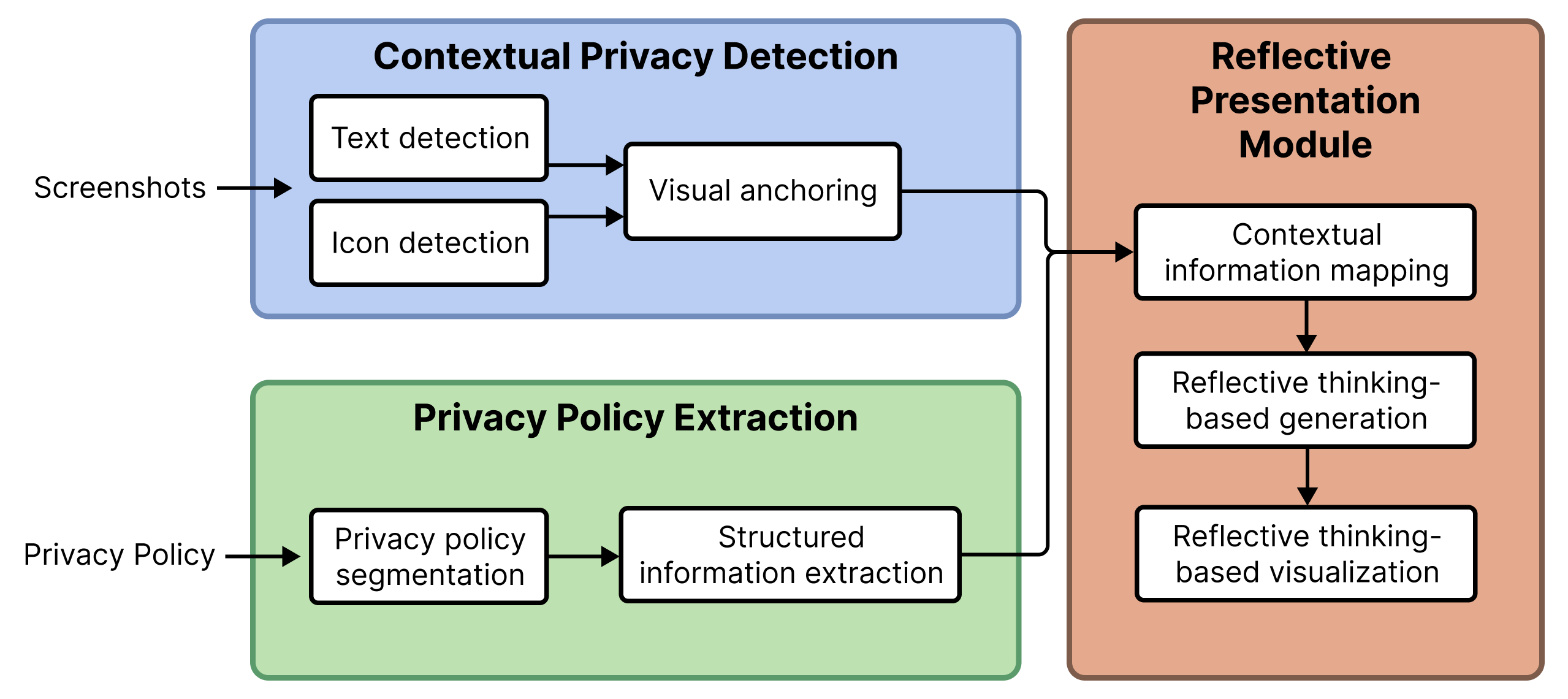}
    \caption{The algorithm flow of \proj{}, which consists of \textit{contextual privacy detection}, \textit{privacy policy extraction} and \textit{reflective presentation module}.}
    \label{fig:system_flow}
\end{figure}

The algorithmic pipeline of \proj{} is composed of three modules, as illustrated in Figure~\ref{fig:system_flow}: (1) \textit{Contextual Privacy Detection}, which identifies sensitive information on the user's screen, (2) \textit{Privacy Policy Extraction}, which retrieves and parses the corresponding data practices from the app's privacy policy, and (3) \textit{Reflective Presentation}, which aligns these two sources of information and generates reflective risk-based notices.

For contextual privacy detection, we adopt a dual-channel pipeline from prior work that processes textual and iconic elements separately~\cite{pan2024new}. Optical Character Recognition (OCR) localizes textual elements, which a LLM classifies into privacy-related data types. Iconic elements are localized as candidate graphical components and classified by a vision model. Their categories are then mapped into the same pre-defined data categories, forming a set of typed contextual elements. The vision model also outlined their visual anchors and later serve as inputs for alignment with the categories of data practices extracted from privacy policies in the reflective presentation stage.

For privacy policy extraction, we located each app's privacy policy by first querying the app's name to search for its policy page on the web (e.g., Google Search). We queried the app's name through locating the name both in the installation page and also in the opening page. We then prompted LLMs to extract the relevant policy text corresponding to each pre-defined data category. For each segment, we prompted LLMs to extract details on data collected, transmission, sharing, disclosure, and other descriptions, based on the extracted privacy policy segment of the previous step.

After contextual privacy detection and privacy policy extraction, \proj{} adopted the reflective thinking framework to present the data practices~\cite{driscoll1994reflective}. It aligns the detected contextual elements with the extracted policy segments by matching their shared, pre-defined data categories. This contextual mapping determines which specifci data practices are relevant to the user's current on-screen activity. Based on this alignment, we then prompted LLMs to aggregate data practices for the specific type of data into a short contextual notice, and also prompted LLMs to augment the aligned segments into reflective descriptions (see Section~\ref{sec:implementation} for the prompt structure). The mapped category, short contextual notice, reflective description and privacy policy segment serve as the basis for interactive presentation to users (see Section~\ref{sec:interaction_design}), where the mapped category serves as the information category displayed in Figure~\ref{fig:teaser_2}, contextual notice and reflective description serves as two notifications (see Figure~\ref{fig:teaser_3}), and the privacy policy segment serves as the excerpt (see Figure~\ref{fig:teaser_4}).

% After contextual privacy detection and the privacy policy extraction, \proj{} adopted the reflective thinking framework to present the data practice~\cite{driscoll1994reflective}. This involved firstly conduct a contextual information mapping to align the contextual privacy and the extracted privacy policy. After this step, we then augmented the extracted structured privacy policy segments as reflective privacy policy descriptions using LLMs as generation models. These risk descriptions aligned with  (see Section~\ref{sec:interaction_design}). 

\subsection{Implementation Details}\label{sec:implementation}

To demonstrate our design's feasibility, we implemented a functional prototype of \proj{}. The contextual privacy detection module, following Pan et al.~\cite{pan2024new}, uses PaddleOCR\footnote{https://github.com/PaddlePaddle/PaddleOCR} and GPT-3.5-Turbo\footnote{https://platform.openai.com/docs/models/gpt-3.5-turbo} for text recognition and a pretrained ResNet~\cite{he2016deep} for icons. For the privacy policy extraction, GPT-4o\footnote{https://platform.openai.com/docs/models/gpt-4o} first retrieves the app's policy text from the web. It then segments the text and extracts structured data practices using a keyword-based mapping adapted from prior work~\cite{pan2024new}, which covers approximately 80\% of common policy topics. Finally, GPT-4o generates reflective risk descriptions for each data category. All LLM tasks utilized a role-playing prompt structure~\cite{shanahan2023role} that (1) defined the role for the LLMs, such as an expert analyzer, (2) defined the task for the LLMs, (3) provided the background knowledge such as the predefined categories for it to refer to, or the original privacy policy, and (4) specified rules for it to adhere to, such as to strictly output the privacy policy content for the segment extraction task, or to generate risks grounded in data practice for the risk generation task. The outputs were parsed before use (see supplementary materials for prompts). To minimize real-time latency, these policy segments, data practices, and risk descriptions are generated and cached upon the app's first launch.

The prototype operates on a client-server architecture, with the mobile frontend transmitting screenshots to a backend server. While custom for this study, this backend could be integrated as a system-level service. To ensure responsiveness, all processing, including LLM reasoning, is parallelized. The system updates its analysis by triggering a new process whenever the pixel-wise overlap between consecutive screenshots falls below an 80\% threshold.

\section{Technical Evaluation of \proj{}'s Components}

We conducted a technical evaluation on different components of \proj{} to prove its technical feasibility, through separately evaluating (1) the accuracy and effectiveness of privacy policy extraction and conversion, and (2) the latency of the contextual privacy detection pipeline, i.e., the time from screenshot arrival to return scenario display. The tests where run on a server with 8~vCPUs and 32~GB RAM.

% The tests except for those calling GPT-4o APIs were run on a Macbook Air with Apple M2 Chip (24GB Memory). 

\subsection{Evaluation of Privacy Policy Extraction and Conversion}\label{sec:evaluation-pp}

We evaluated three parts of privacy policy extraction, including the accuracy of search privacy policy, the accuracy of privacy policy segment extraction, and the usefulness of reflective thinking-based conversion. We used CPP4APP~\cite{pan2024new}, CA4P-483~\cite{zhao2022fine} and MAPP Corpus~\cite{arora2022tale}, which is used in previous work on contextual privacy policy. For MAPP Corpus, we only used the English version privacy policies. We first evaluated the possibility of finding the privacy policy through LLMs. We used LLMs with web search function to search the privacy policy through giving them the name of the APP. The accuracy of finding privacy policies on CPP4APP, CA4P-483 and MAPP Corpus are 100.0\%, 99.2\%, 96.9\% separately, demonstrating our method's feasibility. 

% \begin{table}[!htbp]
%     \centering
%     \caption{Accuracy of finding privacy policies across datasets.}
%     \label{tbl:accuracy_finding}
%     \begin{tabular}{lc}
%         \toprule
%         Dataset & Accuracy (\%) \\
%         \midrule
%         CPP4APP     & 100.0 \\
%         CA4P-483    & 99.2 \\
%         MAPP Corpus & 96.9 \\
%         \bottomrule
%     \end{tabular}
% \end{table}

% \begin{table}[!htbp]
%     \centering
%     \subfloat[The accuracy of finding privacy policies.]{
%     \label{tbl:accuracy_finding}
%     \begin{tabular}{cccc}
%     \toprule
%     Dataset & Accuracy (SD) \\
%     \midrule
%     CPP4APP & $100.0\%$ \\
%     \hline
%     CA4P-483 & $99.2\%$ \\
%     \hline
%     MAPP Corpus & $96.9\%$ \\ 
%     \toprule
%     \end{tabular}
%     }
% \end{table}

We further evaluated the accuracy of privacy policy segment extraction. For each data item we performed three times of evaluation and took the average to avoid random effect. Specifically, we ran the extraction algorithm for each information category and reported the accuracy for each category and overall, as shown in Table~\ref{tbl:accuracy_extraction}. The ratings comprised two parts, where the first part evaluated whether the text was accurately extracted from the original text (i.e., no hallucinations). The second part evaluated whether the text accurately matched the corresponding class (i.e., useful). The background model we benchmarked was GPT-4o, consistent with the model used in our implementation (Section~\ref{sec:implementation}) of \proj{}. We found on average GPT-4o reached over 87\% extraction accuracy, and in specific classes such as contacts and names it reached an accuracy of over 90\% on all datasets, indicating its feasibility.

\begin{table}[!htbp]
    \centering
    \caption{The end-to-end accuracy of privacy policy segments' extraction combining the two parts (i.e., whether the text was accurately extracted from the original text, and whether the text accurately matched the corresponding class).}
    \label{tbl:accuracy_extraction}
    \begin{tabular}{p{2.1cm}p{1.2cm}p{2.0cm}p{1.2cm} p{1.2cm}p{2.0cm}p{1.3cm}p{1.2cm}}
    \toprule
     & \textbf{Location} & \textbf{Address} & \textbf{Phone} & \textbf{Email} & \textbf{Birthday} & \textbf{Contacts} & \textbf{Name} \\
    \midrule
    CPP4APP & $98.0\%$ & $95.9\%$ & $93.8\%$ & $98.0\%$ & $91.8\%$ & $93.9\%$ & $93.9\%$ \\
    \hline
    CA4P-483 & $94.5\%$ & $92.1\%$ & $94.5\%$ & $88.5\%$ & $96.4\%$ & $92.1\%$ & $97.0\%$ \\
    \hline
    MAPP Corpus & $87.5\%$ & $85.9\%$ & $84.4\%$ & $85.9\%$ & $87.5\%$ & $92.2\%$ & $90.6\%$ \\ \midrule
     & \textbf{Voices} & \textbf{Social media} & \textbf{Photos} & \textbf{Profile} & \textbf{Financial info} & \textbf{Avg.} & \\ \midrule
     CPP4APP & $91.8\%$ & $93.9\%$ & $95.9\%$ & $89.8\%$ & $91.8\%$ & $94.0\%$ & \\
    \hline
    CA4P-483 & $95.8\%$ & $95.2\%$ & $89.7\%$ & $93.9\%$ & $89.7\%$ & $93.3\%$ & \\
    \hline
    MAPP Corpus & $87.5\%$ & $84.4\%$ & $95.3\%$ & $84.3\%$ & $86.0\%$ & $87.6\%$ & \\ 
    \bottomrule
    \end{tabular}
\end{table}

We finally evaluated whether the generated risk descriptions are useful, where we recruited 10 participants to each manually label all generated risk descriptions according to the criteria that ``whether the generated risk description would prompt reflections'' on a 7-point Likert scale (1=not at all, 7=very much), following prior guidelines for evaluation~\cite{chen2025clear}. As in Table~\ref{tbl:reflection_risk}, we found on average ratings on all datasets exceed 5.5 out of 7. Specific classes such as \textit{social media} and \textit{contact} reached a higher score of over 6 out of 7, indicating the feasibility of the risk descriptions. 

\begin{table}[!htbp]
    \centering
    \caption{The results of ``whether the generated risk description would prompt reflections'' (1=not at all, 7=very much), where the underscripted number denoted one standard deviation.}
    \label{tbl:reflection_risk}
    \begin{tabular}{p{2.1cm}p{1.1cm}p{2.2cm}p{1.1cm} p{1.1cm}p{2.2cm}p{1.3cm}p{1.1cm}}
    \toprule
     & \textbf{Location} & \textbf{Address} & \textbf{Phone} & \textbf{Email} & \textbf{Birthday} & \textbf{Contacts} & \textbf{Name} \\
    \midrule
    CPP4APP & $6.5_{0.4}$ & $6.6_{0.3}$ & $6.2_{0.3}$ & $6.1_{0.2}$ & $6.6_{0.3}$ & $6.8_{0.1}$ & $5.9_{0.5}$ \\
    \hline
    CA4P-483 & $5.8_{1.0}$ & $5.9_{0.9}$ & $5.4_{0.8}$ & $5.2_{1.2}$ & $6.0_{0.6}$ & $6.2_{0.4}$ & $5.0_{1.2}$ \\
    \hline
    MAPP Corpus & $5.9_{1.1}$ & $5.9_{1.0}$ & $5.4_{0.7}$ & $5.3_{1.3}$ & $5.9_{0.7}$ & $6.0_{0.7}$ & $5.3_{0.8}$ \\ \midrule
     & \textbf{Voices} & \textbf{Social media} & \textbf{Photos} & \textbf{Profile} & \textbf{Financial info} & \textbf{Avg.} & \\ \midrule
     CPP4APP & $6.0_{0.0}$ & $6.5_{0.3}$ & $6.6_{0.3}$ & $6.4_{0.4}$ & $6.5_{0.3}$ & $6.4_{0.4}$ & \\
    \hline
    CA4P-483 & $4.7_{1.4}$ & $6.1_{0.9}$ & $6.2_{0.5}$ & $6.2_{0.5}$ & $6.1_{0.4}$ &  $5.7_{0.9}$ & \\
    \hline
    MAPP Corpus & $5.2_{1.2}$ & $6.1_{0.8}$ & $5.8_{0.6}$ & $5.6_{0.6}$ & $5.5_{0.8}$ & $5.7_{0.9}$ & \\ 
    \bottomrule
    \end{tabular}
\end{table}

\subsection{Evaluation of Contextual Privacy Detection Pipeline Latency}

We evaluated \proj{}'s latency using the CPP4APP dataset~\cite{pan2024new}. We define end-to-end latency as the elapsed time from capturing a screen interface to rendering a CPP to the user. We compared two pipeline configurations: a baseline serial design and our parallelized implementation described in Section~\ref{sec:implementation}, where the GUI-element classification step is executed concurrently across all detected elements on a screen.

% \begin{table}[!htbp]
%     \centering
%     \caption{}
    
% \end{table}

\begin{table}[!htbp]
    \centering
    \caption{Latency of (a) \proj{} (parallel) vs. baseline (serial)  pipelines, and (b) different components in \proj{}. All time are in seconds. SD denoted one standard deviation.}
    \subfloat[Latency of baseline and \proj{}.]{
        \label{tbl:latency_comparison}
        \begin{tabular}{lcccc}
            \toprule
            Pipeline & Avg. & Min & Max & SD \\
            \midrule
            Baseline (Serial) & 19.78 & 3.62 & 137.57 & 14.31 \\
            Conflect (Parallel) & 4.35 & 2.38 & 7.71 & 0.93 \\
            \bottomrule
        \end{tabular}
    }
    \subfloat[Latency of components in \proj{}.]{
        \label{tbl:component_latency}
        \begin{tabular}{lc}
            \toprule
            Category & Latency (SD) \\
            \midrule
            GUI element localization & 2.49 (0.663) \\
            GUI element classification & 1.84 (0.679) \\
            Matching & 0.02 (0.003) \\
            \bottomrule
        \end{tabular}
    }
    % \caption{}
    
\end{table}

Table~\ref{tbl:latency_comparison} summarizes the results, where on average \proj{} achieved a latency of 4.35\,s. Parallelizing the classification stage reduced the average end-to-end latency by nearly 78\%. More importantly, the system became significantly more stable, with the standard deviation reduced by 93\% and the maximum delay cut by about 95\%. These improvements indicate that parallelization not only accelerates the pipeline but also eliminates extreme outliers, resulting in a consistently predictable latency profile. Table~\ref{tbl:component_latency} further breaks down the latency of individual pipeline components. Localization dominates the overall cost, as it combines text detection via PaddleOCR and non-text UI element detection, both executed on a CPU-only server without GPU acceleration. Within this step, text detection alone averages 2.06s, making it the single heaviest subcomponent. In contrast, our classification stage already achieves low latency under parallelization, indicating that the main bottleneck for \proj{} lies in the runtime environment rather than the pipeline design itself.

% First part. Identification accuracy.

% Second part. Matching accuracy.

% Third part. Coverage and time.

% Final part. Contextual warning effectiveness.

% Privacy policy automatic extraction accuracy. (how to evaluate that we could extract privacy policy from installation page and mapping)

% evaluate whether the system could accurately search privacy policy.

% evaluate whether the system could extract privacy policy

% evaluate whether the system could correctly generate contextual privacy policy (this nearly seems end-to-end), including coverage

\section{User Evaluation}

We conducted a user study (N=28) to validate whether \proj{} could help users understand privacy policies without inducing a high cognitive load, and what behavior will users exhibit when facing such CPPs.

\subsection{Participants and Apparatus}

We recruited 28 Chinese participants (16 males, 12 females, $M_{age}=26.7$, $SD=3.3$) through distributing posters on online Wechat groups. 6 have a master's degree or above, 9 have a bachelor's degree and others were with high school's degree or below. 2 are from Computer Science related backgrounds, and 9 are from Engineer backgrounds. Others are from multiple disciplinary. To enhance the ecological validity of our study, consistent with prior practice~\cite{zhang2025privcaptcha}, each participant used their own smartphones to interact with different techniques, which was provided by the researchers for installation. This study got the approval of our Institutional Review Board (IRB), and each participant was compensated 100 RMB for their time. The study (e.g., interface, apps) was conducted in Chinese.

\subsection{Study Design}

We adopted a within-subjects study with a single factor \textbf{\textit{technique}}. This design was chosen to facilitate direct comparisons between techniques while control for individual differences in reading speed and comprehension~\cite{obar2020biggest}. The presentation order of the four techniques was counterbalanced across participants using a Latin-square design to mitigate order effects. Although all participants interacted with the app in the same order, we expected this to have minimal impact on their behavior.

We compared \proj{} with three other techniques (see  Figure~\ref{fig:interfaces}) to span commercial settings and state-of-the-art CPPs~\cite{pan2024new}. Although other privacy awareness tools exist to enhance user awareness of data collection~\cite{stellmacher2022escaping,tabassum2018increasing}, we excluded them as previous research~\cite{reinhardt2021visual} found these alternatives less effective:

\begin{figure}
    \centering
    \subfloat[\proj{}.]{
        \includegraphics[width=0.23\textwidth]{Figure/Conflect_popup.png}
    }
    \subfloat[CPP.]{
        \includegraphics[width=0.23\textwidth]{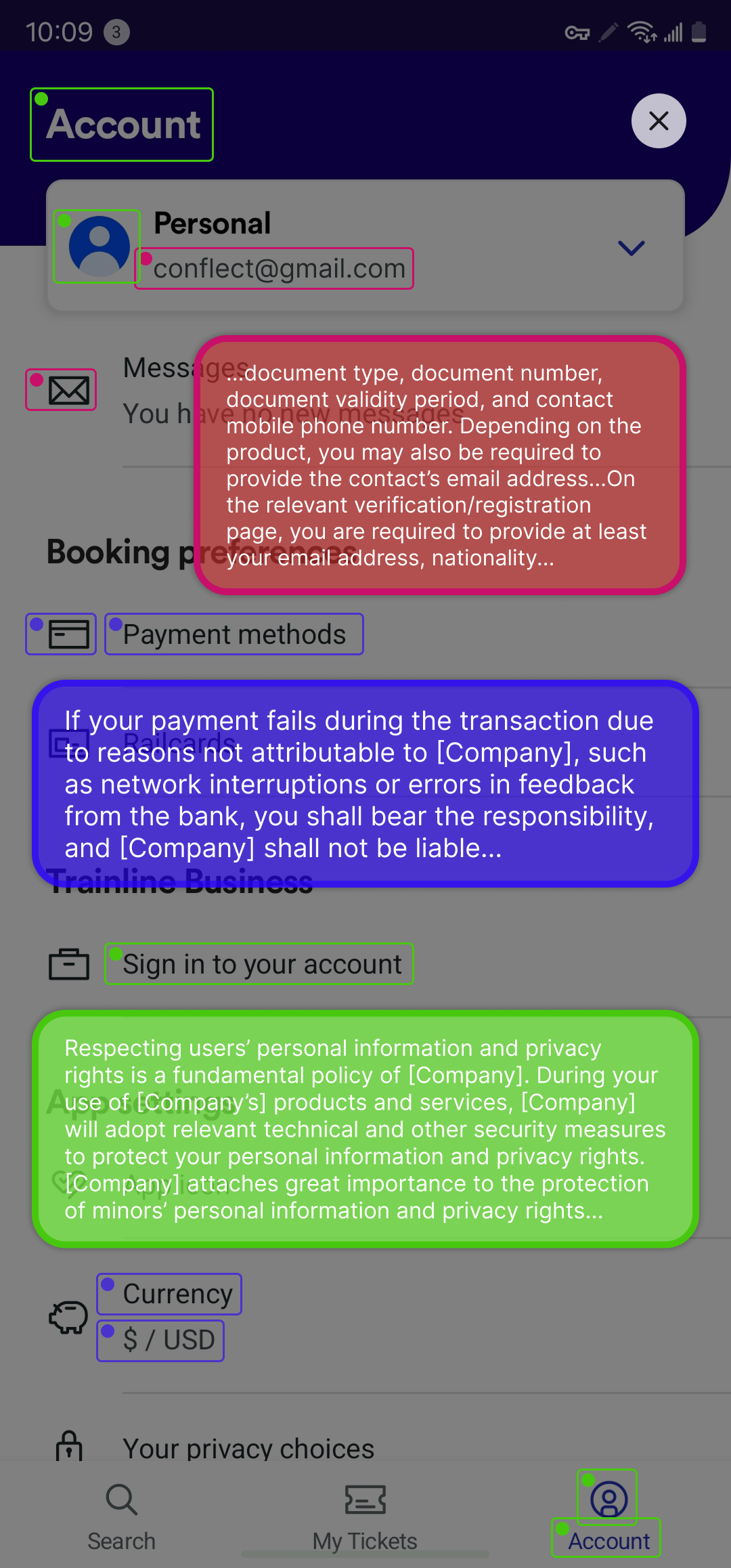}
    }
    \subfloat[Privacy Label.]{
        \includegraphics[width=0.23\textwidth]{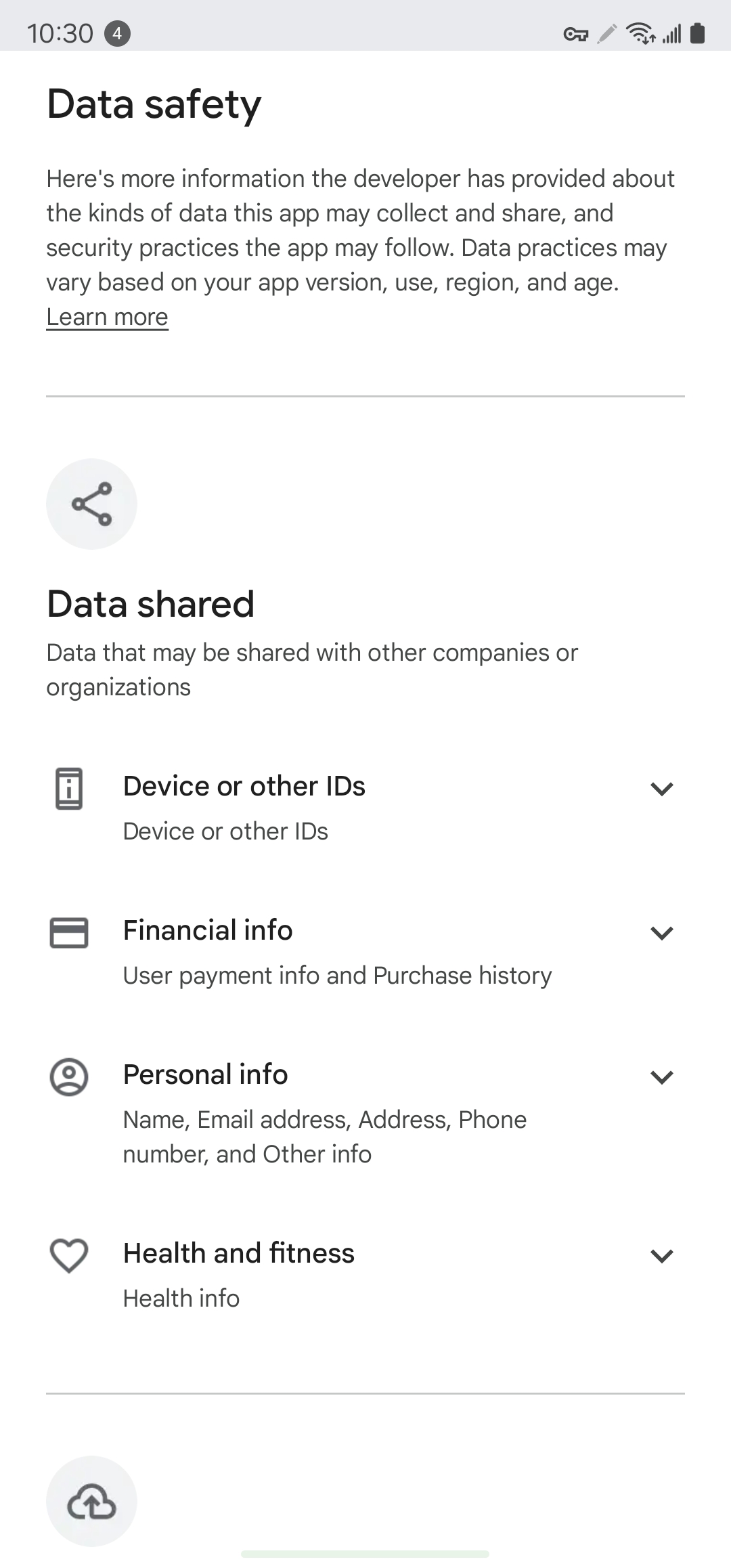}
    }
    \subfloat[Privacy Policy.]{
        \includegraphics[width=0.23\textwidth]{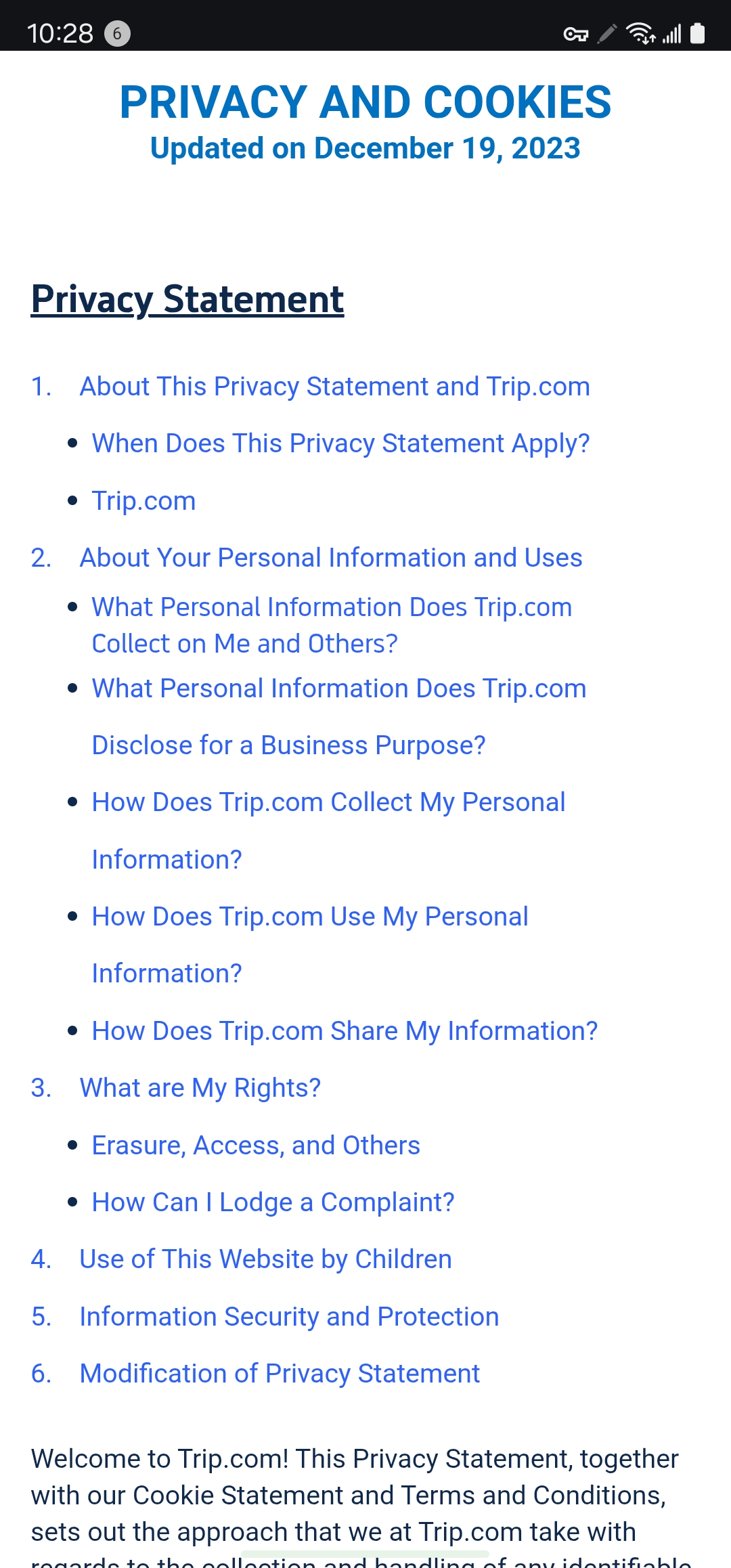}
    }
    \caption{The interfaces of different techniques. The interfaces are originally shown in Chinese and we translated them to English.}
    \label{fig:interfaces}
\end{figure}

$\bullet$ CPP~\cite{pan2024new}: To benchmark \proj{} against the state-of-the-art in CPPs on mobile devices, we replicated the user interface and functionality from a recent work~\cite{pan2024new}.

$\bullet$ Privacy label (PPL)~\cite{kelley2009nutrition,reinhardt2021visual}: This condition represents the prevalent industry standard for summarizing privacy practices~\cite{khandelwal2023comparing}. We used the standard privacy label in the HUAWEI app store to mimic realistic settings.

$\bullet$ Privacy policy (PP): This condition represents the conventional, legally-mandated baseline. To mimic realistic setting, we presented users with the app's original privacy policy, following common practices~\cite{reinhardt2021visual,zhang2025privcaptcha}.

We selected four popular apps across different domains (travel, entertainment, browsing, planning) that users were familiar with but have not read the privacy policies: Ctrip, iQIYI, UC Browser and Tripadvisor. The lengths of their privacy policies were 27,021, 7,619, 8,735 and 24,106 characters respectively. For each app, we designed a realistic, multi-step task. For example, the Ctrip task requires participants to search for a flight, filter the results, select a round-trip ticket and proceed to the payment page by filling in their personal information (see Supplementary materials for the task set). We measured the following aspects to investigate their understanding, experience, trust, and cognitive load:

\textbf{Understanding of privacy policy}~\cite{zhang2025privcaptcha}: We assessed users' comprehension of data practices using a six-question quiz, following validated methods from prior work~\cite{zhang2025privcaptcha,stellmacher2022escaping}. The questions covered key aspects like data collection, usage and sharing. Experimenters manually designed and cross-checked alternative incorrect options through referring to policies in other apps.
% We designed six quiz questions based on prior guidance~\cite{zhang2025privcaptcha}. The questions covered key aspects of data practice, which is emphasized as the most important aspects within privacy policies~\cite{zhang2025privcaptcha,reinhardt2021visual}, including data types, usage, and sharing. We used the same method as prior work~\cite{zhang2025privcaptcha} to reduce bias and validate the questions.

\textbf{System usability and user experience}~\cite{zhang2025privcaptcha,chen2025clear}: assessed using the User Experience Questionnaire Short Version (UEQ-S)~\cite{laugwitz2008construction} and System Usability Scale (SUS)~\cite{brooke1996sus}. 

\textbf{Trust}: we measured the trust in both the app and each technique via the following 7-point Likert scales (1=very unlikely to agree, 7=very likely to agree): ``I trust the original app about its data practice.'', and ``I trust the content \proj{} presented to me.''

\textbf{Cognitive load}~\cite{zhang2025privcaptcha}: measured using NASA-TLX~\cite{hart1988development}. 

The study concluded with a semi-structured interview to gather qualitative insights into the user experience, comprehension, privacy awareness, and decision-making processes for each technique. The full interview script is shown in Appendix~\ref{app:script}.

 % The experiment concluded with semi-structured interview to gather additional insights about users' reflection process. We asked users about the overall experience and satisfaction, their experience around information presentation and comprehension, their privacy awareness and knowledge, their behavior and privacy decision, and finally suggestions for privacy policy visualizations.

\subsection{Procedure}

% During the task, participants needed to view the privacy policy visualization in the corresponding timepoint. For privacy policy and privacy label, they viewed before they opened the app. For the two CPP visualizations, they viewed while using the app. The trial concluded once the participant had completed the task. Since our participants were either native Chinese speakers or highly proficient in Chinese, all interfaces were presented in Chinese to ensure that language barriers did not confound comprehension and engagement. 

After a briefing on the study's objectives, participants provided informed consent. The experiment consisted of four sessions, one for each of the four techniques, and mapped to distinct apps, with the techniques' order counter-balanced and the app's order fixed. In each session, participants used the app to complete the corresponding task. The procedure for interacting with the techniques differed by condition. For PP and PPL, participants reviewed the entire privacy visualization before the task. For \proj{} and CPP, participants interacted with these during the task.

To balance ecological validity with experimental control, we did not impose time limits on the tasks. However, to ensure a fair comparison of comprehension and cognitive load, participants were instructed to review each privacy visualization fully. This protocol was implemented to prevent skimming of lengthy policies, a behavior known to confound comprehension metrics~\cite{zhang2025privcaptcha}. Observations confirmed that all participants adhered to this instruction. A session concluded once the participant completed the assigned task.

\subsection{Analysis Methods}

% For quantitative data, we performed XX analysis and XX analysis. 

% For qualitative interview data, we performed thematic analysis~\cite{} to the transcribed user interviews.
For the understanding of privacy policy, and subjective rating data, as they violated normal distribution, we adopted Friedman non-parametric tests and corresponding Nemenyi post-hoc comparisons. All post-hoc comparisons are performed with appropriate Bonferroni adjustments, and significance are reported with $p < .05$. For qualitative data, we adopted thematic analysis~\cite{braun2021thematic} on the transcribed data, where two primary authors jointly coded the scripts, with intermittent discussions to solve disagreements. 

\subsection{Results}

% Participants interacted with \proj{} with 

\subsubsection{Understanding of Privacy Policy}

\begin{figure}[!htbp]
    \centering
    \includegraphics[width=0.7\textwidth]{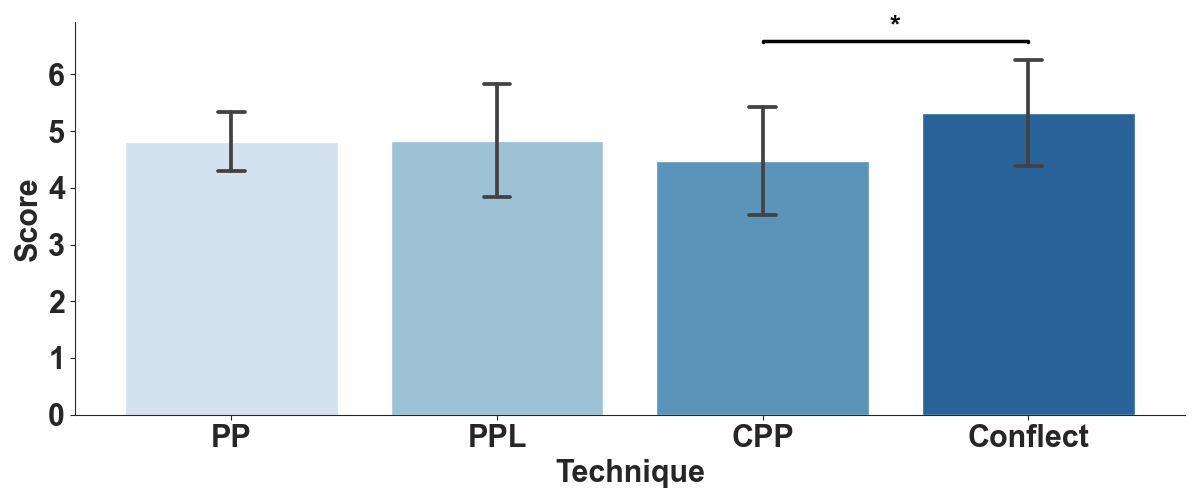}
    \caption{Users' understanding of privacy policies with different privacy policy visualization techniques. Errorbar indicated one standard deviation. * indicated significance at $p < .05$.}
    \label{fig:understanding}
\end{figure}

As shown in Figure~\ref{fig:understanding}, we found a significant effect of technique on users' understandings ($\chi^2_3 = 9.385$, $p = .025 < .05$, Kendall's W = .164), with \textit{\proj{}} reaching significant higher scores than \textit{CPP}. This suggests that \textit{\proj{}} could maintain a comparable or superior capability to let users understand privacy policy.

% \subsubsection{Interaction Statistics}

% Talk about how many times users click on (1) each icon, (2) notification -> privacy policy, (3) ignore / control.
% if no quantitative results -> qualitative

% \begin{figure}[!htbp]
%     \centering
%     \includegraphics[width=0.6\textwidth]{}
%     \caption{Interaction time for privacy policies with different privacy policy visualization technique. Errorbar indicated one standard deviation.}
%     \label{}
% \end{figure}

% \begin{table}[!htbp]
%     \centering
%     \caption{Your Table Caption Here}
%     \label{tab:my_label}
%     \begin{tabular}{|l|c|r|}
%         \hline
%         \textbf{Header 1 (Left-aligned)} & \textbf{Header 2 (Center-aligned)} & \textbf{Header 3 (Right-aligned)} \\
%         \hline
%         Row 1, Col 1 & 100 & \$10.00 \\
%         Row 2, Col 1 & 250 & \$25.50 \\
%         Row 3, Col 1 & 175 & \$17.80 \\
%         \hline
%     \end{tabular}
% \end{table}

\subsubsection{Subjective Ratings}

\begin{figure}[!htbp]
    \subfloat[SUS.]{
        \includegraphics[width=0.35\textwidth]{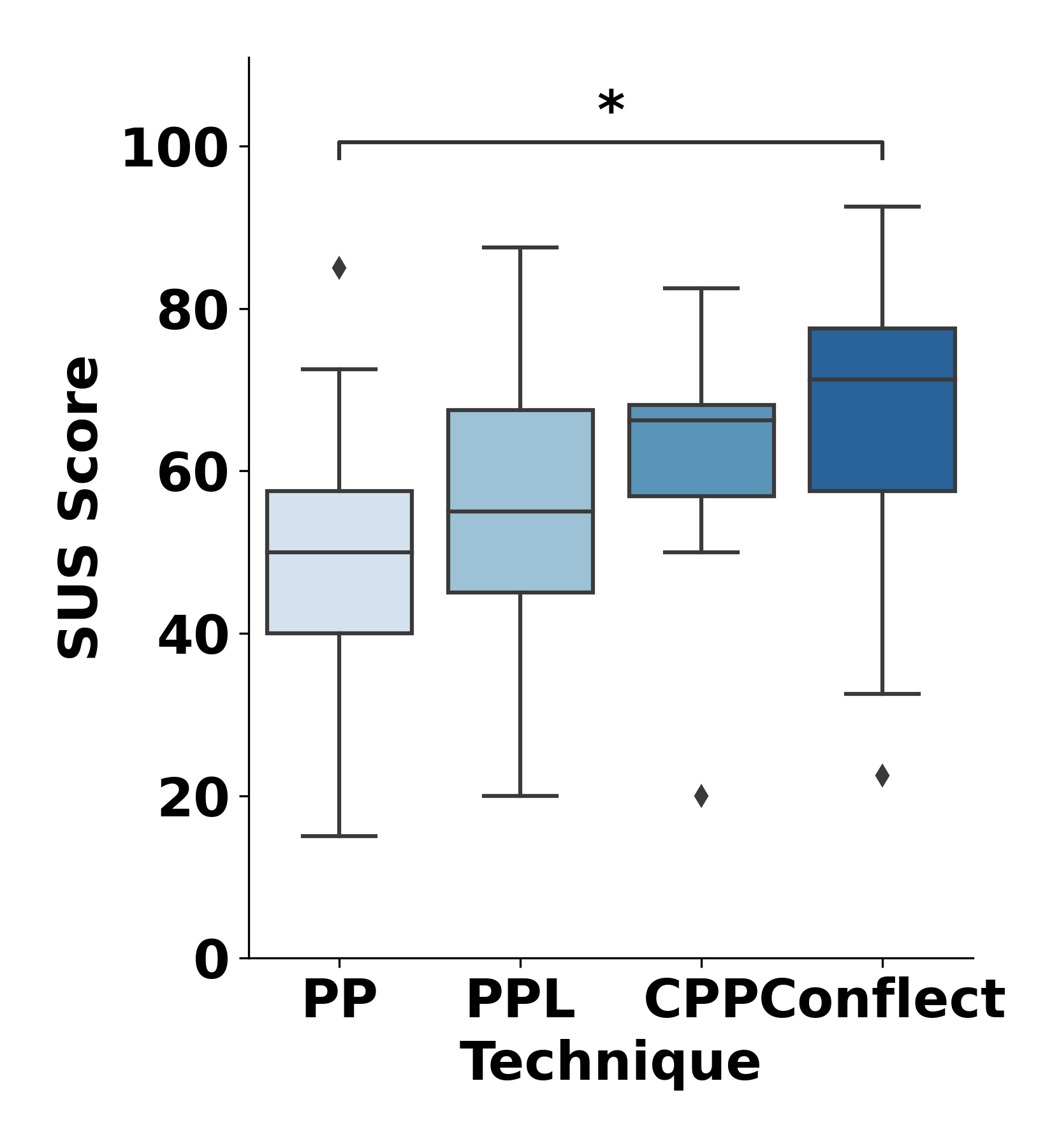}
        \label{fig:sus_scores}
    }
    \subfloat[Trust.]{
        \includegraphics[width=0.4\textwidth]{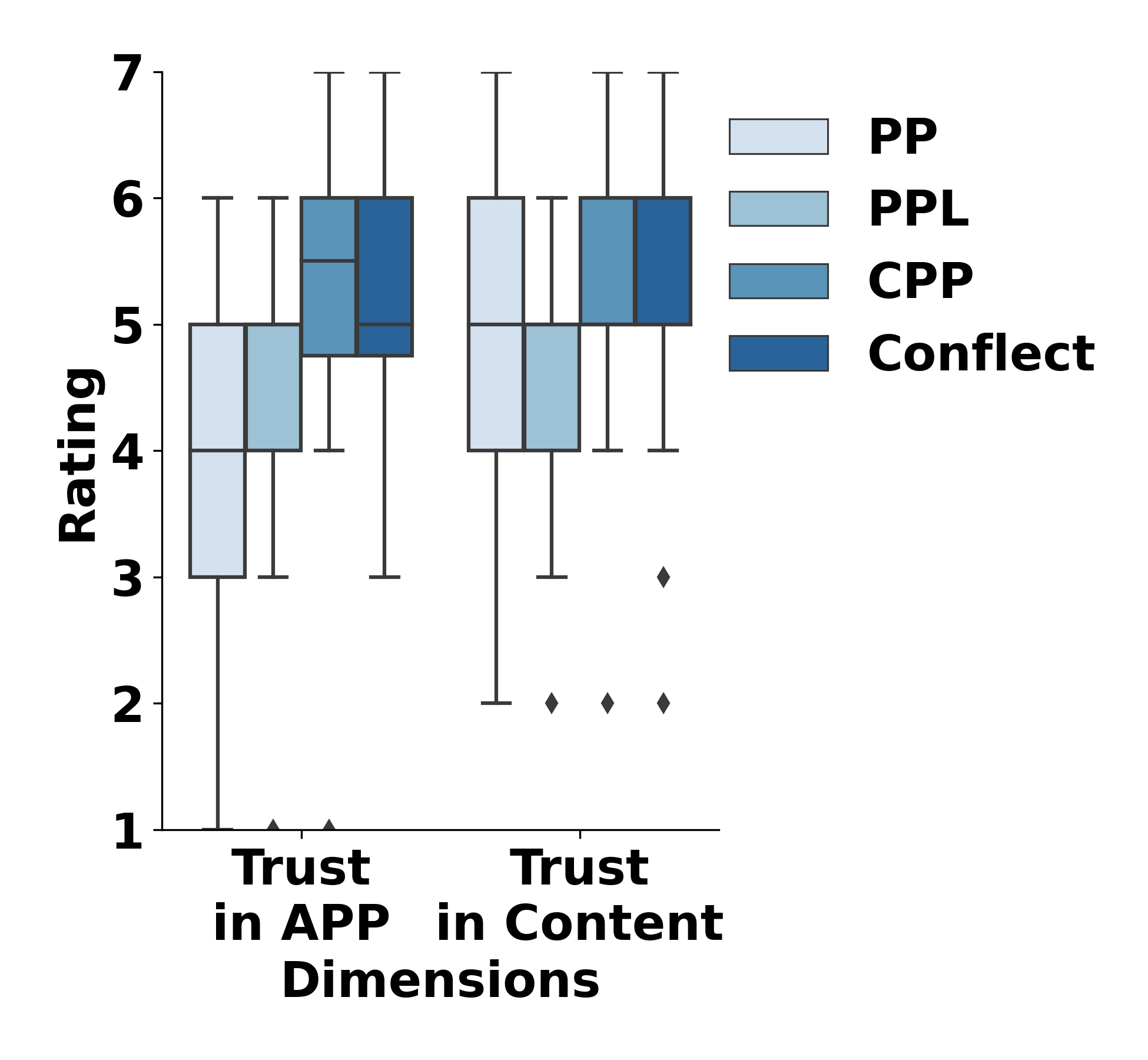}
        \label{fig:trust_scores}
    }
    \caption{The (a) SUS and (b) trust score across different techniques (1: most negative, 7: most positive). * indicated significance at $p < .05$.}
    \label{fig:sus_trust}
\end{figure}

Our analysis of subjective ratings reveals that \textit{\proj{}} significantly enhances the user experience, particularly in terms of usability and cognitive load, when compared to \textit{privacy policy}. Regarding system usability (see Figure~\ref{fig:sus_scores}), we found a significant effect of the technique on SUS ratings ($\chi^2_3 = 10.473$, $p = .015$, Kendall's W = $.205$). Post-hoc tests confirmed that \textit{\proj{}} was perceived as significantly more usable than the traditional privacy policy ($p = .028 < .05$). In terms of trust (see Figure~\ref{fig:trust_scores}), while there was a significant main effect on users' trust in the app ($\chi^2_3 = 11.290$, $p = .010 < .05$, Kendall's W = $.221$), post-hoc comparisons did not reveal significant differences between specific techniques. We also found significant effect on users' trust in the presented content ($\chi^2_3 = 8.698$, $p = .034 < .05$, Kendall's W = $.171$), but with no post-hoc significance. This suggests that while \textit{\proj{}}'s interactive design improves usability, its current primary impact is on the user experience rather than altering fundamental trust in the application itself.

% Figure~\ref{fig:sus_scores} showed the ratings for trust. We found a significant effect of technique on users' trust in app ($H = 8.698$, $p = .034 < .05$, $\eta^2_H = .091$), however we did not found significant difference in post-hoc comparisons. We did not find a significant effect of technique on users' trust in content ($H = 4.609$, $p = .20$, $\eta^2_H = .026$). Figure~\ref{fig:trust_scores} showed the ratings for SUS. We found a significant effect of technique on SUS ratings ($H = 8.767$, $p = .03$, $\eta^2_H = .093$), with Dunn's post-hoc test finding significant differences between \textit{\proj{}} and \textit{privacy policy} ($p = .036 < .05$).

\begin{figure}[!htbp]
    \includegraphics[width=0.75\textwidth]{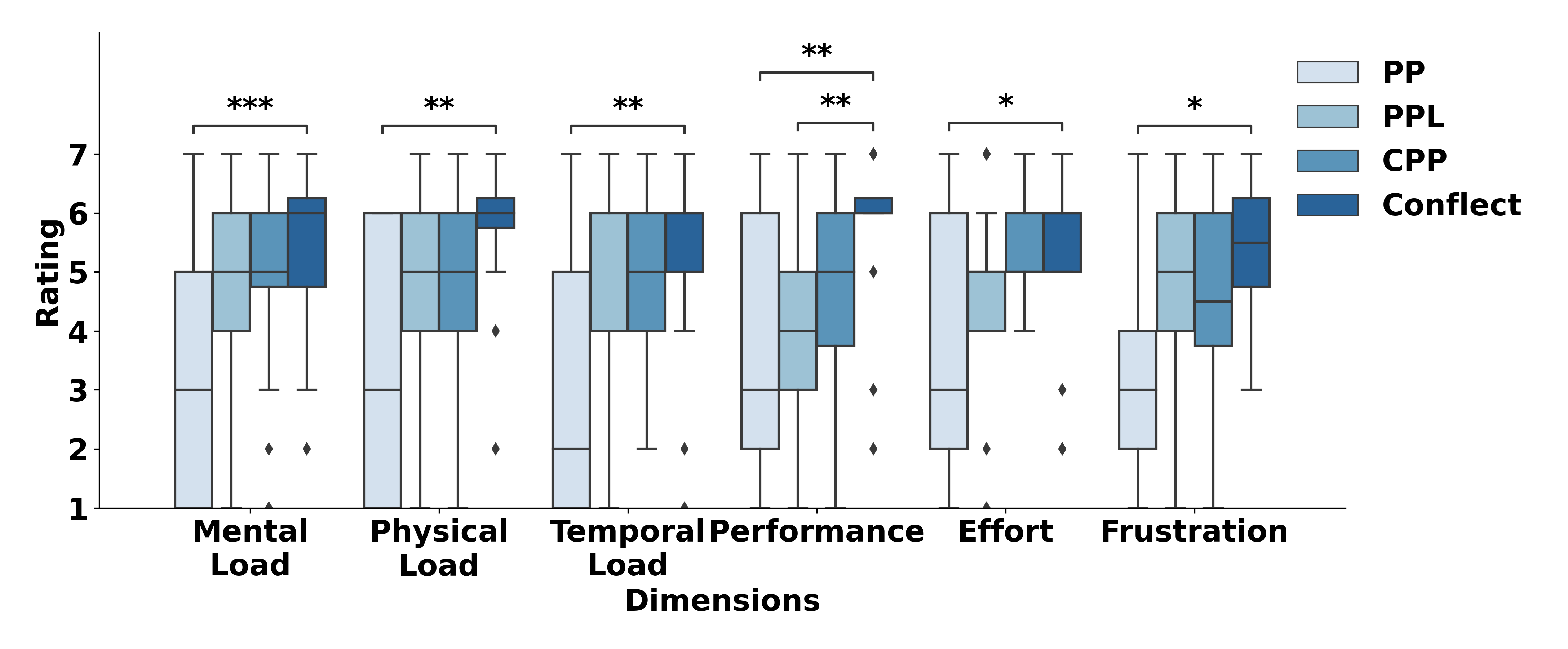}
    \caption{NASA-TLX ratings across techniques, all dimensions reverse-coded, except performance retained in its original direction; in all cases, higher rating indicates more positive outcomes. *, **, *** indicated significance at $p < .05$, $p < .01$, $p < .001$ respectively.}
    \label{fig:nasa_tlx}
\end{figure}

A central objective of \textit{\proj{}} was to reduce the cognitive burden of comprehending privacy information, and the NASA-TLX results affirm its effectiveness (see Figure~\ref{fig:nasa_tlx}). We found that \textit{\proj{}} significantly outperformed the traditional privacy policy across all six dimensions of cognitive load. Specifically, participants reported significantly better mental load ($\chi^2_3 = 16.490$, $p = .0009 < .001$, Kendall's W = $.323$), physical load ($\chi^2_3 = 15.490$, $p = .001 < .01$, Kendall's W = $.304$), temporal load ($\chi^2_3 = 15.826$, $p = .001 < .01$, Kendall's W = $.315$), performance ($\chi^2_3 = 13.961$, $p = .003 < .01$, Kendall's W = $.274$), effort ($\chi^2_3 = 9.415$, $p = .024 < .05$, Kendall's W = $.184$) and frustration ($\chi^2_3 = 9.686$, $p = .021 < .05$, Kendall's W = $.194$). Post-hoc comparisons found significant differences between \textit{\proj{}} and \textit{privacy policy} (mental load: $p = .001 < .01$, physical load: $p = .003 < .01$, temporal load: $p = .003 < .01$, performance: $p = .007 < .01$, effort: $p = .049 < .05$, frustration: $p = .011 < .05$). These findings collectively indicate that \textit{\proj{}} makes the process of engaging with privacy policies feel substantially less demanding and more efficient, reducing user frustration.

The UEQ ratings further illuminate the specific advantages of \textit{\proj{}}'s design. \textit{\proj{}} was rated as significantly more perspicuous than both the \textit{privacy policy} and \textit{privacy label} ($\chi^2_3 = 16.981$, $p = .0007 < .001$, Kendall's W = $.335$, post-hoc \textit{\proj{}} vs. \textit{privacy policy}: $p = .006 < .01$, post-hoc \textit{\proj{}} vs. \textit{privacy label}: $p = .006 < .01$). It was also seen as more efficient ($\chi^2_3 = 9.628$, $p = .022 < .05$, Kendall's W = $.188$, post-hoc $p = .034 < .05$) and dependable ($\chi^2_3 = 21.300$, $p = .0001 < .001$, Kendall's W = $.420$, post-hoc $p = .0003 < .001$) than the \textit{privacy policy}. This suggests that users found it easier to achieve their goals with a clear and reliable interface. 

Furthermore, \textit{\proj{}} successfully made the experience more engaging. It was rated as significantly more interesting ($\chi^2_3 = 25.617$, $p = .0000 < .001$, Kendall's W = $.502$, post-hoc \textit{\proj{}} vs. \textit{privacy policy}: $p = .0004 < .001$, post-hoc \textit{\proj{}} vs. \textit{privacy label}: $p = .006 < .01$) and novel ($\chi^2_3 = 19.096$, $p = .0002 < .001$, Kendall's W = $.345$, post-hoc \textit{\proj{}} vs. \textit{privacy policy}: $p = .001 < .01$, post-hoc \textit{\proj{}} vs. \textit{privacy label}: $p = .016 < .05$) than both the \textit{privacy policy} and the \textit{privacy label}. Participants also found \textit{\proj{}} significantly more comfortable ($\chi^2_3 = 15.978$, $p = .001 < .01$, Kendall's W = $.313$, post-hoc \textit{\proj{}} vs. \textit{privacy policy}: $p = .003 < .01$) and leading-edge ($\chi^2_3 = 26.500$, $p = .0000 < .001$, Kendall's W = $.520$, post-hoc \textit{\proj{}} vs. \textit{privacy policy}: $p = .0000 < .001$, \textit{\proj{}} vs. \textit{privacy label}: $p = .001 < .01$) than the baselines. This demonstrates that \textit{\proj{}} not only improves the functional aspects of policy comprehension but also enhances the affective experience, making users more willing to engage with important privacy information.

\begin{figure}[!htbp]
    \includegraphics[width=\textwidth]{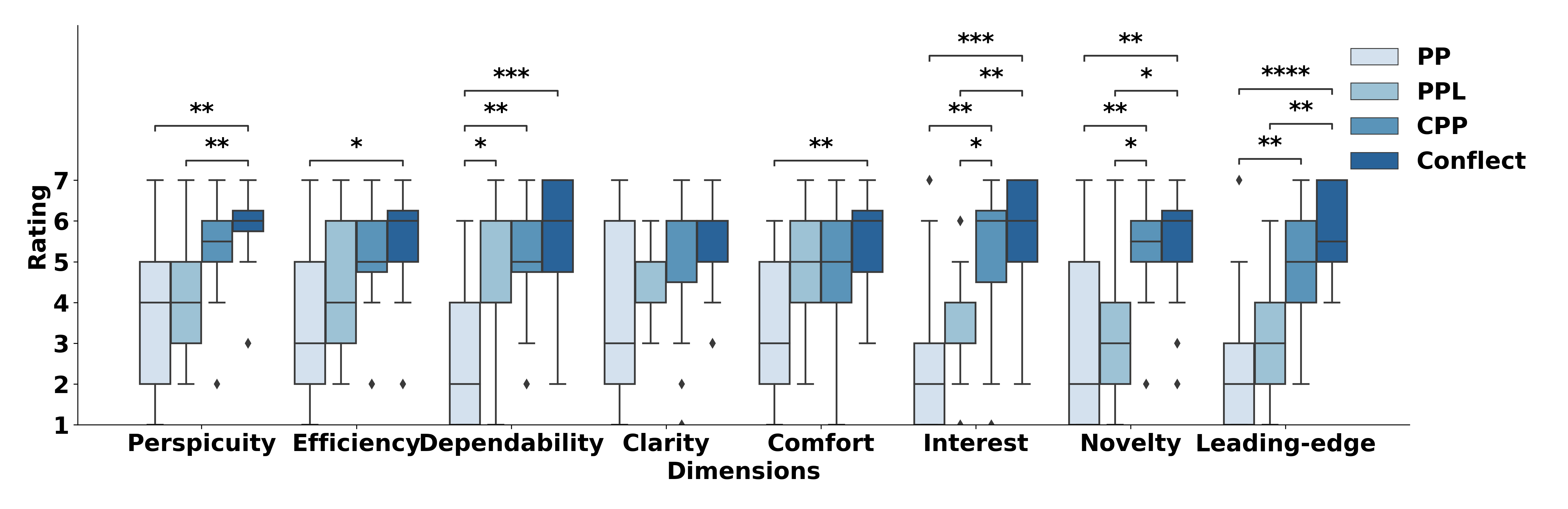}
    \caption{UEQ ratings across different techniques (1: most negative, 7: most positive). Errorbar indicated one standard deviation. *, **, ***, **** indicated significance at $p < .05$, $p < .01$, $p < .001$, $p < .0001$ respectively.}
    \label{fig:ueq}
\end{figure}

\subsubsection{Subjective Feedback}\label{sec:feedback}

% What we cares: (1) behavioral, how they use (include behavioral, timing) -> behavioral change, (2) trust stances, (3) any potential improvements.

\textbf{Overall experience and initial impressions} Participants' overall experience with \proj{} was predominantly positive, especially when contrasted with traditional privacy policies and more intrusive CPP implementations. The technology was frequently described as ``convenient'', ``concise'', and a significant improvement over reading lengthy, text-heavy documents. Participant P10 summarized this sentiment, stating, \textit{``It's very convenient, unlike the [other] CPP which displays a large amount of information''}. Another user, P2, found reading standard policies ``head-hurting'' and described the return to a tool like \proj{} as ``very comfortable'' and ``easy to understand''.

The most consistently praised feature was the tool's ability to provide real-time, context-specific privacy information without significantly disrupting the user's primary task. Participant P18 articulated this key benefit: \textit{``I think it can concisely convey information to me in the form of an icon... It quietly reminds me there without disturbing me, and when I want to check it, it can provide me with information. I think this is what impressed me the most''}. Similarly, P6 noted that the tool caused ``almost no interference'' and that the entire task progression felt ``relatively smooth''.

However, the experience was not universally seamless. A minority of users reported lag for pop-ups and interface update. These technical imperfects colored participants' overall perception despite appreciating the concept.

\textbf{User interface and interaction design.} Feedback on \proj{}'s UI/UX centered on the floating sidebar and its associated icons and notifications. The visual design of the icons was a notable strength. Multiple participants found them intuitive and effective at conveying the type of data being collected at a glance. P6 praised them as ``quite stunning'' and having ``high recognizability'', allowing them to \textit{``realize what kind of private information is being accessed here''} even without clicking. This pre-emptive understanding was seen as a key usability win, as it reduced the time needed to process the information upon expansion. The use of color-coding, such as red, yellow and green, to signify risk levels was also identified as an impressive and memorable feature.

The unobtrusive nature of the floating widget was a primary point of positive comparison against more aggressive, full-screen CPP. P4 found the latter ``a bit troublesome'' as it required an extra click to dismiss before continuing a task, and P18 echoed that \proj{} was a quiet, persistent reminder that did not demand immediate interaction. 

\textbf{Information presentation and comprehension.} \proj{}'s method of abstracting and presenting privacy information was highly valued by participants. The consensus was that summarizing lengthy policies and generating potential risks via AI was far more effective than presenting raw policy text.

P18 articulated this value clearly: \textit{``It is far more helpful to me than simply displaying the privacy terms, because I would not want to spend time reading those privacy terms. I would rather have something to summarize and simplify it for me, give me a reminder, and let me focus on the key points''}. P14 agreed, noting that without the AI-generated summaries, they might be confused by professional terminology and \textit{``wouldn't spend extra time looking up the real meaning of the term''}. Participants found the provided summaries to be ``very easy to understand'' and ``more colloquial'' than the dense language of the full policies. Design elements like highlighting key terms were also appreciated for drawing attention to the most critical information.

\textbf{Impact on privacy awareness and knowledge.} A primary success of \proj{} was its ability to enhance users' awareness and understanding of data collection practices. A recurring theme was the discovery of ``hidden'' or unexpected data collection events. Several participants learned that apps collect data far more frequently than they had previously assumed. P6 expressed surprise, stating, \textit{``I used to think that... it would only request information after I complete some operations... but it also collects specific information when I click on places where I obviously don't need to input content''}. This sentiment was shared by P13, who realized that apps do not always ask for permission for each instance of data collection and that \textit{``it may already be collecting my information when I click a certain button''}. For these users, \proj{} provided a novel and sometimes unsettling look into the continuous nature of background data access.

The tool also helped users understand the purpose behind certain data requests, which in some cases reduced suspicion. P17 noted, \textit{``After its explanation, I felt it was indeed necessary... to give me recommendations''}. This indicates the tool can build understanding, not just raise alarms. For many, this was a significant educational experience. As P14 noted, they learned that they could proactively disable permissions, a fact they were previously unaware of, assuming an app's requests were ``a matter of course''.

\textbf{Influence on user behavior and decision-making.} While \proj{} was effective at raising awareness, its direct impact on in-the-moment user behavior was more limited. The majority of participants acknowledged that the tool's prompts did not cause them to alter their actions during the experiment. P18 explained this by citing the task-oriented nature of the scenario: \textit{``a large part of it is because I have a task to complete... In a real-life scenario, I might still prioritize completing my task, with privacy being a secondary issue''}. P17 gave a similar reason, stating their focus was on \textit{``following the task step by step''}. P15 attributed their lack of behavioral change to a generally low personal concern for privacy.

However, the tool was not without influence. Some users reported that the information prompted increased deliberation and caution. P2 stated that after seeing a warning about email marketing, they would \textit{``reconsider and maybe provide partial information instead of all of it''}. P14, upon reading a policy summary, decided to alter their real personal information to ``randomly typed data'' to avoid having their true details collected, a direct change in behavior motivated by the tool. This suggests that while it may not always stop a user's workflow, \proj{} can encourage more privacy-conscious choices when users perceive a direct risk.

\textbf{Trust and adoption.} The question of whether participants would use \proj{} long-term was intrinsically linked to trust. Most participants expressed a high degree of trust and willingness to use the tool if it were integrated into the mobile operating system by the manufacturer, such as Apple, Android and Xiaomi. P2 stated they would have \textit{``90\% trust''} if it were system-native. P18 provided a sophisticated rationale based on accountability: \textit{``When the software has a problem, who is responsible for it?... If it is integrated into Xiaomi, then I know you have been recognized by Xiaomi. When it has a problem, I can not only look for you but also for Xiaomi.''}

\textbf{Potential improvements for future design and customization.} Participants provided feedback for future iterations, with a strong emphasis on icon design, personalization and control.

\textit{Icon design.} Despite the general praise, a few users found the icons' meanings were not immediately clear. P13 reported needing to \textit{``spend some time understanding the meaning of each symbol''} and stated that without reading the instructions, \textit{``I wouldn't know what it means''}. P17 echoed this, noting, \textit{``I couldn't quite understand those icons... I had to click on them to see the specific introduction''}. This suggests an initial learning curve for the iconography.

\textit{Functional and UI enhancements.} Users also proposed new functionalities. P18 wished for an ``abnormality alert'' that would flag when an app's data collection practices deviate from industry norms, providing a more powerful warning for outlier behavior. P14 suggested a feature that would allow \proj{} to directly ``control permissions'' and block an app's frequent requests on the user's behalf. P10 wanted the tool to go beyond just flagging risks and actively guide them on how to \textit{``avoid privacy risks for the next step''}

\textit{Customization and personalization.} Users expressed a desire to tailor the tool's notifications to their individual privacy preferences. P15 suggested a system for \textit{``customizing the level of privacy''}, where they could designate specific data types like ``payment account information'' as highly sensitive, which would then trigger a more prominent alert, such as a different color. P13 echoed this, wanting strong alerts for \textit{``very private information like my address, ID number, or phone number''}, while suggesting that alerts for less sensitive data like account information could be ``weaker''. This points to a need for a user-defined risk model rather than a one-size-fits-all approach.

\section{Discussions}

% We first discussed the feasibility of \proj{}, then discussed its control and the future envision of contextual privacy policy in particular.

\subsection{Feasibility of \proj{}}

\textbf{Accuracy.} While \proj{} demonstrated superior usability (Figure~\ref{fig:sus_trust}) and 94.0\% accuracy on CPP4APP (Table~\ref{tbl:accuracy_extraction}), similar to prior systems~\cite{pan2024new,zhang2025privcaptcha}, it still has failing cases. These cases can be categorized into two classes. The first class contained those mistakenly detecting non-sensitive elements as sensitive ones. While this might seem like a drawback, it can function as a time for users to learn about privacy policies, similar to sandbox-based approaches~\cite{chen2024empathy}. The second class contains those mistakenly detect sensitive elements as non-sensitive ones. This is more important, as genuine risks may be overlooked. This could be mitigated by the fact that participants' scrolling or clicking triggers re-detection, which complements the previous detection results. 

Despite these limitations, user feedback indicates that \proj{} could act as a useful complement to existing privacy policies (Section~\ref{sec:feedback}). It functions as a system service to alert users to potential risks, especially in contexts where app developers are not legally required to disclose all data collection practices. Nonetheless, we acknowledge that an imperfect accuracy could potentially lead to user mistrust or, conversely, over-trust. This issue could be partially mitigated by providing clear and appropriate notifications to manage user expectations~\cite{ahn2024impact}.

\textbf{Trust and behavioral change.} Trust is a critical factor influencing CPP adoption. Our findings indicate trust levels depend heavily on the deployment model of \proj{}. Participants reported the highest trust when \proj{} was framed as a native system service, citing confidence in their smartphone manufacturer. In contrast, a standalone third-party app elicited mixed reactions: some were open, but others doubted its motives and saw the screen-recording permissions as intrusive without pre-existing trust. Integrating \proj{} directly into the host app raised the most suspicion, as participants found it paradoxical for an app to both collect data and warn users about its own practices.
% Trust is a critical factor influencing the adoption of CPPs. Our findings indicate that trust levels are highly dependent on the deployment model of \proj{}. Participants expressed the highest degree of trust when \proj{} was conceptualized as a native system service, attributing this to the inherent confidence they place in their smartphone's manufacturer. Conversely, deploying \proj{} as a standalone third-party app elicited mixed reactions. While some participants were open to it, others expressed skepticism regarding the app's motives and perceived the requisite screen recording permissions as intrusive, particularly without pre-existing trust in the app. Notably, integrating \proj{} directly within the primary application was viewed the most suspicion. Participants found it paradoxical for an app to simultaneously collect data while warning users about its own practices. 

Although a minority of participants appreciated the transparency of a company ``selflessly'' disclosing its data practices, translating this awareness into behavioral change proved challenging. While some expressed a willingness to switch to alternative applications, others desire the risk notifications to appear earlier to prevent them from using this app. This revealed a paradox that users desire early warnings, yet the most effective and actionable notifications are inherently contextual at the moment of data collection. This tension suggests that future work could explore sandbox-based implementations~\cite{chen2024empathy}, which might allow users to experience potential data risks in a realistic yet safe environment before engaging with an application.
% Participants consistently confronted a privacy-utility trade-off, often prioritizing immediate task completion over privacy concerns. 

\textbf{Deployment.} To transition \proj{} from a functional prototype to a practical system service, especially granting users the control, it must have real-time access to an application's data collection practices. This requires two primary capabilities. First, it must be able to ascertain an application's declared permissions, which can be achieved by querying the operation system's permission manager via APIs. Second, it needs to identify user-provided data, necessitating the monitoring of user interface interactions or user touching to capture data inputs as they occur. Even so, a system-level CPP may not reliably intercept all data flows. Sensitive information entered transmitted via third-party SDKs often bypasses OS-level checkpoints, which is hard for permission system or on-screen monitoring to intervene. These require LLMs to dynamically interpret risks based on the temporal operational sequences~\cite{shaoprivacylens}. 
%Furthermore, a temporal mismatch may still exist between when a risk is detected from screen context and when the actual data disclosure occurs, although it is significantly improved compared with traditional privacy policies and privacy labels (Figure~\ref{fig:sus_trust}). For example, a financial risk warning might appear while a user is merely browsing a payment screen, long before or after the transaction is executed, making timely control difficult. 

\textbf{Latency.} While demonstrating the feasibility of near real-time CPPs, with users generally reporting the interaction as usable (see Section~\ref{sec:feedback}), \proj{}'s average end-to-end latency of 4.35 seconds remains a limitation for just-in-time interventions. In practice, this delay could cause notifications to lag behind user interactions, undermining their contextual relevance. Achieving true real-time performance needs further optimization. One promising avenue include leveraging GPUs for processing policies and images, which could reduce inference time to sub-second range for tasks like OCR\footnote{\url{https://paddlepaddle.github.io/PaddleOCR/main/en/version2.x/legacy/benchmark.html}}. Besides, developing lightweight, domain-specific detection models and integrating \proj{} at system level to utilize rich signals such as API calls or keystroke events may reduce processing and enable effective interventions. 

\subsection{Usefulness of Control Options and The Related Privacy-Utility Trade-Off}\label{sec:discussion-control}

Consistent with prior work on CPPs~\cite{pan2024new,windl2022automating,zhang2025privcaptcha}, we prioritize user comprehension and reflection over technically implementing direct ``just-in-time'' controls. This aligns with reflective thinking frameworks, which treat awareness and understanding as prerequisites for meaningful action~\cite{Murmann2021FromDR}. While technically feasible to implement controls, we found a fundamental privacy-utility trade-off in users' control. 

Granular controls frequently compel users to make a difficult choice between restricting data access and maintaining application functionality, leading to users' control fatigue~\cite{choi2018role}. This dilemma is a well-documented tension where convenience often overrides privacy concerns, a pattern also observed in human-LLM interactions~\cite{zhang2024s}. Although they sometimes could choose alternative apps, most of the time they have no choices, especially when an application's function is perceived as irreplaceable, effectively removing the user's agency. Therefore, we position CPPs to serve a dual role. Beyond simply informing users, they can function as a contextual auditing tool~\cite{pan2024new,chen2025clear}, making the data practices of apps transparent, and thereby prompting app service providers to change the app's practices. 

\subsection{Generalizability of \proj{}}

We scoped CPPs for Android apps (see Figure~\ref{fig:interfaces}) and deliberately excluded apps that operate primarily as background services, such as malware scanners and junk cleaners, as they lack the interactive elements necessary to trigger contextual notices.

While our implementation is Android-specific, the foundational principles of CPPs are potentially generalizable. It could be translated to iOS ecosystems through translating these codes to swift language. Besides, our methods could theoretically be adapted for websites on mobile phones where the site's name, and interface icons are discernible~\cite{windl2022automating,ortloff2020implementation}. ``Microapps'', such as those within the WeChat ecosystem, represent another potential application. Although these third-party services have limited permissions, they still pose privacy risks when users input sensitive data. While beyond the current focus of \proj{}, our framework's core reflective thinking-based design could be extended to these interfaces. However, extending this to desktop websites would demand significant efforts, probably following the methods by Windl et al.~\cite{windl2022automating}.

Conceptually, CPPs are distinct from runtime permission notices~\cite{schaub2015design}. CPPs are argued to align better with users' mental models of data privacy because they are triggered by user interaction with specific interface elements that imply data collection, rather than by underlying code execution~\cite{li2024we,pan2024new}. \proj{} helps bridge the gap where an interface may suggest data collection without an explicit permission request, or when a permission request lacks clear visual context, where \proj{}'s control could be integrated as runtime notices. By linking notices directly to user-perceived actions, we envision CPPs to complement runtime controls, offering clear explanations and intuitive choices.

\section{Limitations and Future Work}

We acknowledge that our paper has two limitations. First, our paper focused on the design of CPPs, while acknowledging that certain parts of privacy policies may not be covered by CPPs~\cite{pan2024new}, and in the current stage CPPs mainly act as a complementary module of privacy policies. A possible mitigation could place these non-contextual information at the installation time~\cite{zhang2025privcaptcha,reinhardt2021visual}. Second, our recruitment may subject to bias, which involved primarily of young Chinese participants. As privacy is cultural sensitive and documented by the prior work~\cite{xu2024dipa2}, Chinese users may focus on different information than those from other cultures, therefore the effectiveness of our design on contextual privacy policy in other cultures warrants further examination. Nevertheless, most of our participants are studying and working in different parts of the world, which may have broadened their perspectives beyond their cultural background. 

\section{Conclusion}

This paper explores how to improve user understanding and engagement through designing CPPs with a reflective thinking-based framework. We first conducted three workshops with experienced designers and researchers, eliciting six problems and three design dimensions to solve these problems. The most critical issue was the disconnect between context and action. Our findings led to the design of \proj{}, a prototype with contextual detection, policy reasoning and matching modules. We instantiated \proj{} as a functional prototype, where a technical evaluation showed \proj{} achieved 94.0\% accuracy on the CPP4APP dataset, with an end-to-end latency of 4.35 seconds. A user study with 28 participants demonstrated that \proj{} significantly enhances user understanding, reduces cognitive load, and increases satisfaction compared with privacy policies, privacy labels and CPPs.

% Privacy policies are lengthy and users also have no motivations to read those privacy policies. Contextual privacy policy, which segments the privacy policies and present them in a contextual manner could potentially increase users' understanding, but how to motivate users and balance their cognitive load warrants future examination. This paper starts with three workshops with experienced designers and researchers to elicit design ideas of contextual privacy policy guided by the reflective thinking framework. Participants articulated 20 design instances, and 6 problems. Through a convergence phase, Participants thought attracting users' motivation, and balancing tasks and privacy policy reading are the most important aspects. Participants also articulated that a floating-window based reminder and a notification based design as the best design considering effectiveness and feasibility. We designed and implemented the most prominent design as \proj{}, composing of contextual detection modules, privacy policy reasoning modules and matching modules. Through a technical evaluation, \proj{} achieved 94.0\% accuracy for extracting privacy policies on CPP4APP dataset, and an end-to-end latency of 4.35\,s on a server with 8~vCPUs and 32~GB RAM. The user study (N=28) proved \proj{}'s effectiveness in facilitating users' understanding, lowering users' cognitive load and increasing users' satisfaction in comparison to traditional privacy policy, naive contextual privacy policy, and privacy notice techniques. 

\section{Acknowledgments of the Use of AI}

We used AI, in particular LLMs, in the backend of \proj{}. 
Specifically, GPT-3.5-turbo was employed to classify detected text after OCR-based text detection (Section~\ref{sec:implementation}). 
GPT-4o was used in multiple parts of our system: (1) to retrieve, segment, and structure privacy policy text and to generate reflective, risk-based scenarios from extracted policy segments (Section~\ref{sec:implementation}); 
and (2) as the background model in our technical evaluation, where it was benchmarked for accuracy of policy extraction and category-level classification (Section~\ref{sec:evaluation-pp}). 
The authors take full responsibility for the content generated and its integration into this work.

\bibliographystyle{ACM-Reference-Format}
\bibliography{sample-base}

\appendix 

\section{Ethics Considerations}

We acknowledged that our paper may have ethic concerns, and followed Menlo report~\cite{bailey2012menlo} and Belmont report~\cite{beauchamp2008belmont} in designing the studies. Our studies received the approval of our institution's Institutional Review Board (IRB). Before each study, including the workshops, the ones in the technical evaluation, and the final user study, we provided participants with comprehensive information regarding the study's purpose, potential risks, and harms. Participants were explicitly informed of their right to withdraw from the experiment at any time and to have their data removed. We obtained informed consent from every participant.

Each participant was appropriately compensated for their time and efforts. To protect their privacy, all collected participants' data were anonymized and securely stored on our institution's local server.

Our research aims to improve CPPs, thereby empowering individuals to make informed decisions about their privacy, which is a societal benefit. Following the studies, we encouraged participants to actively take steps to protect their personal privacy.

\section{Interview Script For The Evaluation Study}\label{app:script}

A. Overall Experience

$\bullet$ Could you please describe your overall experience using [Tool Name]? What aspects stood out to you the most?

$\bullet$ How easy or difficult did you find the tool to use and understand? Did the presence of the tool interfere with or disrupt your normal workflow in any way? Please elaborate.

$\bullet$ During your use of the tool, were there any specific interactions or prompts that you found particularly memorable?

B. Information Presentation and Comprehension

$\bullet$ What are your thoughts on the way the tool presented privacy information (e.g., in the sidebar, using bullet points, highlighting keywords)? How clear and effective was this presentation?

$\bullet$ Regarding the content, how easy or difficult was it for you to understand the summaries extracted from the full privacy policy and the descriptions of potential risks? Did you encounter any jargon or technical terms that were unclear?

$\bullet$ The tool presents both excerpts from the privacy policy and AI-generated summaries of potential privacy risks. In your opinion, what is the value of this combined approach? Did you find it more helpful than seeing only the original policy text?

C. Privacy Awareness and Knowledge Enhancement

$\bullet$ Before using this tool, how would you describe your level of awareness regarding how this app collects and uses your personal data? What were your primary privacy concerns at that time, if any?

$\bullet$ After using the tool, has your understanding of data privacy changed in any way? Specifically, what new information or insights did you gain?

$\bullet$ Were there any data handling practices or potential risks highlighted by the tool that you were previously unaware of?

D. Behavioral Impact and Decision-Making

$\bullet$ Can you recall any specific instances where the information provided by the tool influenced an action or a decision you were about to make?

$\bullet$ Were there any situations where you chose to ignore a prompt from the tool and proceed with an action? If you're comfortable sharing, could you describe your thought process at that moment?

$\bullet$ What is your opinion on the value of displaying the ``ignore" button?

E. Future Design and Extension

$\bullet$ If you were the designer of this tool, what is one feature you would most want to improve or add to make it more useful for you?

$\bullet$ What are your thoughts on having the ability to customize the tool's settings, such as the content of its prompts, the way information is displayed, or when the prompts are triggered?

$\bullet$ Looking beyond this study, could you see yourself using a tool like this in your daily life? Why or why not?

$\bullet$ How would you describe your level of trust in the tool itself? Would that trust be different if the tool were a native feature provided by your phone's operating system?

\end{document}